\documentstyle[epsf,dcolumn]{mn}
\begin{document}
\newcolumntype{d}[1]{D{.}{.}{#1}}
\newcommand{\twiddles}{\sim}
\newcommand{\etal}{{et al}\/.}
\newcommand{\uv}{{\it uv}}
\def\Ssin#1C#2 {#1C\,#2}
\def\Ss#1{\Ssin#1 }
\def\Ssf#1{\Ssin#1 }
\def\Ssq#1{}
\def\afterpage#1{}
\def\dgr{^\circ}
\title[FR\,II radio galaxies with $z < 0.3$ -- I]{FR\,II radio galaxies
with $\bmath{z < 0.3}$ -- I. Properties of jets, cores and hot spots}
\author[M.J.~Hardcastle \etal]{M.J.~Hardcastle, P.~Alexander,
G.G.~Pooley and J.M.~Riley\\Mullard Radio Astronomy Observatory, Cavendish
Laboratory, Madingley Road, Cambridge, CB3 0HE}
{\par
 \begingroup
\twocolumn[
\vspace*{17pt}\raggedright \sloppy\huge\bf {FR\,II radio galaxies
with $\bmath{z < 0.3}$ -- I. Properties of jets, cores and hot spots}\vskip 23pt\LARGE
M.J.~Hardcastle$^{1,2}$\footnotemark, P.~Alexander$^{1}$,
G.G.~Pooley$^1$ and J.M.~Riley$^1$\par
\small{\it $^1$ Mullard Radio Astronomy Observatory, Cavendish
Laboratory, Madingley Road, Cambridge, CB3 0HE\\
$^2$ Department of Physics, University of Bristol, Royal Fort, Tyndall Avenue, Bristol BS8 1TL}\par
\vskip 22pt
\today\par\vskip 22pt
]
 \thispagestyle{titlepage}
 \endgroup
}
\begin{abstract}
In previous papers we have discussed high-resolution observations of a
large sample of powerful radio galaxies with $z<0.3$. Jets are
detected in up to 80 per cent of the sample, and radio cores in nearly all
the objects; in addition, we are able to resolve the hot spots in most
sources. In this paper we present measurements of the radio properties
of these components.

The prominences of the jets detected do not appear to be a
function of radio luminosity, providing the clearest evidence yet that
the reported low detection rate of jets in radio galaxies has been an
artefact of low-sensitivity observations. We find a positive
correlation between the total source length and core prominence in the
narrow-line radio galaxies. We have found evidence for a relationship
between hot spot size and total source size, but few other significant
relationships between hot spot properties and those of the jets or
lobes. We compare our measurements to those of Bridle \etal\ (1994),
based on observations of a sample of quasars, and argue that the
results are consistent with a modification of the unified model in
which the broad-line radio galaxies are the low-luminosity
counterparts of quasars, although the situation is complicated by
contamination with low-excitation radio galaxies which appear to have
radio properties different from the high-excitation objects. We
discuss the classes of empirical model that can be fitted to the
dataset.
\end{abstract}

\footnotetext{Present address. E-mail {\it M.Hardcastle}@{\it bristol.ac.uk}}

\section{Introduction}

Detailed images of classical double radio sources are essential if we
are to understand the physics of such objects.  In an earlier paper
(Hardcastle \etal\ 1997: hereafter H97) we discussed high-resolution
imaging of a sample of FRII (Fanaroff \& Riley 1974) radio galaxies
with $0.15 < z <0.3$ drawn from the complete sample of Laing, Riley \&
Longair (1983: hereafter LRL). In this paper we combine this sample
with the objects with $z<0.15$ described in Black \etal\ (1992:
hereafter B92) and Leahy \etal\ (1997: hereafter L97) to form a sample
of 50 objects. High-resolution observations are available for almost
all this combined sample, as described in B92, L97 and H97, and jets
are detected in up to 80 per cent of the objects. In these earlier
papers the properties of individual sources were discussed in
detail. In the current paper we present and analyse systematic
measurements made from these maps, particularly of the components
(jets and cores) which directly relate to the energy transport in the
sources, and investigate a number of trends and correlations in the
data.

\subsection{Unified models at $\bmath{z< 0.3}$}
\label{unif}

Unified models for classical double radio sources, in which the FRII
radio galaxies are the parent population of radio-loud quasars (Scheuer
1987; Barthel 1987, 1989) are now widely accepted. The differences
between the radio structures of radio galaxies and quasars are
explained in terms of relativistic beaming of components of the
sources (see section \ref{relbeam}), while anisotropic obscuration
explains the optical differences between them. Barthel (1989) showed
that in the redshift range $0.5 < z < 1$ the relative numbers of
quasars and radio galaxies in LRL, and their distributions of linear
sizes, were consistent with every radio galaxy in the sample being a
misaligned quasar.

However, a problem arises when applying the simplest version of the
unified model to the sources at $z<0.3$ considered in this paper.
There are no FRII quasars in 3CR with $z<0.15$ (Spinrad \etal\ 1985)
or in LRL with $0.15<z<0.3$. These samples are selected on the basis
of low-frequency flux density and so are thought to be free of
orientation bias. If unified models are correct, the absence of
quasars from the sample discussed here implies the presence in it of
other objects aligned at a small angle to the line of sight which
should be detectable by anisotropic optical and radio
emission. Barthel (1989) suggested that the broad-line radio galaxies
(BLRG) were intermediate in viewing angle between quasars and the more
common narrow-line radio galaxies (NLRG), which may be true in some
cases; but in this sample at least the complete absence of
lobe-dominated quasars makes it seem more likely that some or all of
the objects classed as BLRG are true quasars whose optical continuum
is insufficiently bright for them to be classed as such
optically.\footnote{It is important to realise that classifications in
terms of broad or narrow lines are dependent on high-quality spectra,
which are not in general available for this sample. Laing \etal\
(1994) have shown that the classifications can change significantly
with improved observations. The classifications used here must be
viewed as best guesses only; in general they agree with those of other
workers, e.g.\ Jackson \& Rawlings (1997).} In this paper we shall test
this model by comparing the radio properties of the BLRG in our sample
to those of the NLRG and of powerful quasars.

The situation is rendered still more complicated by the presence in
the sample of a number of low-excitation radio galaxies (LERG). This
class of object was first discussed in Hine \& Longair (1979); here we
use the definition of Laing \etal\ (1994: hereafter L94). It is
suggested (e.g.\ Barthel 1994) that these objects should not show
broad emission lines whatever their angle to the line of sight, and
that they form the parent population of BL Lac objects rather than
core-dominated quasars; they will then certainly confuse any attempt
to analyse orientation effects from radio data unless they are treated
separately. If this is done carefully they should provide a
valuable population for comparison with BLRG and NLRG if no
other effects are present. However, we shall show that other effects
do appear to be present in the sample discussed here.

In what follows we shall therefore discuss the overall properties of
the sample with the emission-line classifications of the sources in
mind. We shall compare the properties of this sample with those of
others, particularly the high-resolution images of quasars in the
similar study of Bridle \etal\ (1994; hereafter B94); the place of
these objects in unified models should be borne in mind.

In the present sample there are 15 LERG, 9 BLRG and 25 NLRG. If the
low-excitation objects are discarded, as discussed above, the
proportion of BLRG to NLRG is consistent with the critical angle of
40--$50\dgr$ of Barthel (1989).

\subsection{Relativistic beaming and radio source components}
\label{relbeam}

The emission from any component of the radio source moving at a
significant fraction of the speed of light in the galaxy rest frame
will be anisotropic. As seen on Earth, a feature with velocity $v = \beta c$ and Lorentz factor
$\gamma = (1-\beta^2)^{-{1\over 2}}$ has a flux density $S_{obs}$ which can be
related to the flux density it would have had at rest ($S_{rest}$) by the
formula

\begin{equation}
S_{obs} = S_{rest}[\gamma(1 - \beta \cos
\theta)]^{-(m+\alpha)}\label{db}
\end{equation}
(Ryle and Longair 1967) where $\theta$ is the angle made by the
velocity vector with the line
of sight, $\alpha$ is the spectral index (defined throughout in the
sense $S \propto \nu^{-\alpha}$), and $m$ is a constant
reflecting the geometry of the beamed component -- $m=2$ will be used
for jets.

As discussed above, relativistic beaming is often used as an
explanation for the observed one-sidedness of jets, and for the
greater prominence of jets and cores in quasars as compared to
NLRG. Measurements of superluminal motion in the parsec-scale jets of
quasars and BLRG (corresponding to the cores seen on maps of the
large-scale radio structure) have established that relativistic
velocities are present there. In analysing the present sample we shall
compare the properties of features which might be supposed to be
beamed, with the aim of establishing the degree to which relativistic
velocities persist on large scales.

\subsection{Definitions and conventions}

In what follows we shall make extensive use of terms describing the
features of radio sources, and so some definitions are
important. We use the traditional term `core' for the component,
normally unresolved on arcsecond scales and having a flat spectrum,
which coincides with the central regions of the optical host galaxy.
B94 use the term `central feature' for these components. A `jet' is a
feature conforming to the definition of Bridle \& Perley (1984); it is
at least four times as long as it is wide, separable at high
resolution from other extended structure, and aligned with the core
where it is closest to it. `Possible jets' are features that are not
four times as long as they are wide (they may be knots or trains of
knots), or that have not been imaged at high resolution, but which
meet the other criteria for a jet and are plausible tracers of the
`beam', the underlying stream of particles. We generally use the term
`jet' for the brighter jet in the source (or the only jet if only one
is detected) and the term `counterjet' for the fainter jet where one
is present. `Hot spots' are structures associated with the termination
of the beam. We follow the definition of L97 and H97, who define
them as features that are not part of a jet and that have a largest
dimension smaller than ten per cent of the main axis of the source, a
peak brightness greater than ten times the r.m.s. noise, and a
separation from nearby peaks by a minimum falling to two thirds or
less of the brightness of the fainter peak. In this paper we shall be
interested in the most compact or `primary' hot spots, which seem
likely to be the regions where the beam terminates.

Values of $H_0 = 50$ km s$^{-1}$ Mpc$^{-1}$, $q_0 =0$ are used
throughout this paper.

\section{Analysis and results}

\subsection{The sample}

The source information for the combined
sample is shown in Table \ref{combsamp}.

B92 selected 3CR sources with $P_{178} > 1.5 \times 10^{25}$ W
Hz$^{-1}$ sr$^{-1}$ and $z<0.15$. They excluded sources known not to
have hot spots (wide-angle tail, fat-double and cluster-centre
objects) and three giant radio galaxies. In addition, they did not
image the sources included by LRL from outside the 3CR catalogue;
DA\,240 and \Ss{4C73.08} would have met their selection criteria but
could be excluded as giants in any case. Conversely, almost all the
sources selected by B92 but not LRL are only excluded from LRL's
sample on the grounds of position on the sky [the exceptions are
\Ss{3C197.1}, \Ss{3C223.1} and \Ss{3C277.3} which have $S_{178} <
10.9$ Jy on the Baars \etal\ (1977) scale]. All the sources in the
LRL-based sample defined in H97 meet the power criterion used by B92,
and sources from 3CR are only excluded on the basis of position on the
sky or $S_{178} < 10.9$ Jy. Thus, although the combined sample is not
strictly flux-complete, we do not believe it to be biased in any way
which should affect our analysis, and a complete sub-sample consisting
of the non-giant LRL sources (34 in total) can be constructed if
completeness is an issue; the objects drawn from this sample are
marked in bold type in Table \ref{combsamp}. Possibly more serious
than the power and flux density constraints from the point of view of
bias is the exclusion from our LRL-based sample of sources classed by
them as FRIs. This excludes some objects which are in other respects
similar to those in our sample; for example, \Ss{3C288} (Bridle \etal\
1989) and \Ss{3C346} (Spencer \etal\ 1991) are excluded, though they
are very similar in appearance to \Ss{3C438} or \Ss{3C15}. Generally
this bias manifests itself as an exclusion of sources that have
structure intermediate between classical FRIs and FRIIs, and so does
not affect conclusions based on the sample of sources with classical
FRII structure (the most obvious indicator of this being well-defined
hot spots).

178-MHz flux densities were taken from LRL where possible, and otherwise
determined following the prescription in LRL, correcting to the
scale of Baars \etal\ (1977). The 178-MHz flux densities and low-frequency
spectral indices of the sources in the sample of B92 are discussed in
L97.

The sample of quasars observed by B94, which we use for comparison
with our results, consists of 13 quasars from LRL with angular sizes
greater than 10 arcsec. The redshifts of these objects range from
0.311 to 2.012, with the median redshift being 0.77. Because of the
selection criteria, the sample is biased towards larger linear sizes
with respect to LRL quasars as a whole. The biasing and the inclusion
of sources at very much larger redshifts and luminosities than those
of our sample mean that the B94 sample is not ideal for such
comparisons; however, it is the only sample for which information on
quasar jets is presently available.

\subsection{Analysis of radio maps}

High-resolution, well-sampled electronic radio maps were available to
us for 44 of the 50 sources. The remaining six objects were omitted
essentially randomly, and so do not bias the discussion. From these
images we have measured various parameters which relate to energy
transport in these sources. These are discussed below.

With the exceptions noted in individual tables, the flux densities given are
taken from VLA maps, and are therefore subject to errors of a few per
cent because of the limiting accuracy of the absolute calibration of
the VLA. This error has not been included in the errors quoted.

For ease of tabulation, we have divided the jets, lobes and hot spots
of each source into `north' and `south'.

\subsubsection{Total and lobe flux densities}

A reliable measurement of the total flux density of a source could be
obtained from direct integration on low-resolution maps only if the
object had been observed with sufficient short-baseline coverage to
sample all the large-scale structure. This was true of most objects
observed in H97, but a number of the objects from the sample of B92
were undersampled due to insufficient observation or large angular
size. In these cases (see Table \ref{numbers}) integrated flux
densities are taken from the literature or estimated by interpolation
of the integrated spectrum. Reliable flux densities of individual
lobes are generally only available for the well-sampled sources (Table
\ref{lobes}), though in some cases a zero-spacing flux density from
the literature was used to constrain the short baselines in
maximum-entropy mapping routines.  It was
sometimes difficult to separate the emission from the two lobes; this
is indicated by a correspondingly large error in Table \ref{lobes}.
Largest angular sizes of the sources were measured directly from the
maps, and the distance from the core to the most distant region of
lobe emission is also tabulated so as to define a measure of hot spot
recession comparable with that of B94 -- in the few `winged' sources
where the greatest distance or the largest angular size was in a
direction transverse to the source axis, this is noted and an
alternative distance given.

\subsubsection{Cores}

In general the radio cores of the objects in the sample were
unresolved with the highest resolution of the VLA. The cores were
therefore well fitted by an elliptical Gaussian plus baseline (using
the AIPS task {\sc jmfit}). The integrated flux densities determined
by these fits are given in Table \ref{numbers}; errors are the formal
$1\sigma$ errors from the fitting routine.

\subsubsection{Jets and counterjets}

It is possible, as discussed in B94, to measure the flux density of
jets in a number of different ways. Clearly a simple measurement of
the flux density of the jet region (as in B92) is not useful, even if
the jet region is well defined, because of the necessity of correcting
for background emission (particularly important in these objects where
the jets have low contrast with the lobes). In addition, in a number
of cases the jet region is not well defined, particularly where the
jet is entering a hot spot.

To allow a comparison with the results of B94, we determine jet flux
density using their methods. The total jet flux density is obtained by
integrating over the area that clearly contains jet emission,
correcting for the background by integrating a similarly-shaped area
on either side and subtracting the average of these measurements
normalised to the area of the jet integration. We repeated this
process three times for each jet and for counterjets where
present. Errors are assigned by considering the standard deviation of
the integrated and the normalising values, and combining the resulting
errors in quadrature. Where there is a substantial difference between
the background flux densities on the two sides of the jet, this will
dominate the error assigned.

Where even the most tentative jet or counterjet candidate is detected,
we have chosen to record its flux density. (Flux densities of
`possible' jets are marked with an asterisk in Tables \ref{jet_tab}
and \ref{stjet_tab}.)
Where there was no detection of any sort, we have tried to determine
an upper limit on the flux density of a present but undetected
jet. There is no way in which a truly conservative upper limit can be
set on flux density from an unseen jet, given the number of possible
forms that jets can take, but some reasonable approximations can be
made to a plausible limit that most unseen jets will not exceed. For
compatibility with the measurements of B94 we defined total jet limits
in a similar way. On maps of intermediate resolution we integrated
over the rectangle two beamwidths wide between the core and the
primary hot spot (chosen as the most plausible path for a jet) or the
closest approach which did not intersect bright confusing lobe
emission, and over the two rectangles on either side of it. If the
flux density in the central rectangle was larger than that in the
other two, we subtracted the mean flux density of the edge rectangles
from that of the central one and used the result as an upper
limit. Otherwise, we used the positive difference between the central
flux density and the lower of the two edge flux densities as the upper
limit. This way of assessing the upper limits has the advantages that
the number obtained increases with the inhomogeneity of the lobe (the
correct behaviour, since jets are hidden by confusing lobe structure
as much as by on-source noise), and that when measured on a lobe that
{\it does} have a jet the upper limit is a reasonable measure of the
jet's true flux density.  The obvious disadvantage of this method is
that it is resolution- and sensitivity-dependent, and that a
subjective decision must be taken to make the measurements from a
particular map; where only low-resolution maps were available to us,
the limits are higher.

\label{strjet}
We also measured straight jet flux density. As B94 point out, FRII
radio sources often have a relatively straight inner section followed
by a bent region. Regions of bending are likely to be the sites of
strong interaction between the jet and its environment, which
introduces a further source of random scatter into the measurements;
it is also impossible to apply a simple relativistic beaming analysis
(section \ref{relbeam}) to a bent jet, since there is not a single
value of the angle made by the beam to the line of sight. The straight
jet region is defined, following B94, to be a rectangle, aligned with
the core, over which the line of the brighter or only jet deviates
from the centerline of the rectangle by less than a jet radius and
which avoids any significant confusion with lobe emission. (Thus, when
the jet enters a hot spot or region of high surface brightness, the
straight jet is considered to end.) The rectangle was taken from the
core to the furthest extent of the straight jet, even where the
straight jet was only visible in part of the lobe, to maximise the
chances of a counterjet detection. Identically-sized rectangles on
either side of the straight jet region were integrated to provide an
estimate of the background flux density. The similar rectangular
region on the other side of the core was integrated to provide an
estimate of the flux density from the counterjet, if present, or an
estimated limit on its flux density if not (if the central value was
higher than the values on either side, their average was subtracted;
if not, the positive difference between the lower value and the
central value was used as an upper limit). As before, three
measurements were made on each source and averages taken; errors were
assigned in the same way as for the total jet flux densities. The
straight jet fluxes are tabulated in Table \ref{stjet_tab}.  Most of
the objects with well-defined counterjets have an angle of not quite
$180\dgr$ between jet and counterjet; counterjet measurements of this
sort can thus underestimate the counterjet flux density. On the other
hand, the counterjet rectangle can intersect confusing emission in the
counterjet lobe, and therefore seriously overestimate the counterjet
flux density. These results should therefore be treated with some
caution. Where there was no candidate straight jet in either lobe, no
entry is made in Table \ref{stjet_tab}. In these cases the upper
limits on total jet flux, discussed above, may be taken as upper
limits on straight jet flux.

All the errors on the jet flux density determinations are conservative,
because of the difficulty in deciding which parts of the object should
be treated as a jet or counterjet. In a few cases interpretation is
crucial; these are now discussed.

\begin{itemize}
\item \Ss{3C192}: we have only considered the jet seen near the S hot spot
complex in the high resolution maps of L97. The inclusion of the faint
linear feature seen in the S lobe at low resolution -- which is not
clearly aligned either with the core or with the hot spots -- would
increase the jet flux density substantially.
\item \Ss{3C234}: we have only considered the jet leading into the
northern hot spot. If the ridge E1 of H97 were included (as it is in the
straight jet flux density determination) the integrated jet flux density would be
substantially higher.
\item \Ss{3C300}: It is not clear how much of this object's
north lobe should be classed as a jet. We omitted the last few
arcseconds where there is no surrounding lobe emission. The flux
density might
be increased by up to 15 mJy if this region were included.
\item \Ss{3C327}: we have differed from L97 in counting the faint linear
features in both lobes as possible jets. If only the jet candidate
around the core is considered, the flux density is approximately 0.4 mJy.
\item \Ss{3C390.3}: we only consider material closer to the core than knot B of
Leahy \& Perley (1995), on the grounds that knot B is the primary hot
spot.
\item \Ss{3C403}: we only consider the regions F7, F8 of B92 as true jet,
on the grounds that F6 is the primary hot spot. Including F6 and the
jet-like components F4 and F5 would increase the flux density to 80 mJy.
\item \Ss{3C424}: we have not considered the northern `jet' of this object
to be a true jet. It is however quite possible that some of the
internal structure in this northern component should be treated as jet
material. We have treated S3 as the only jet component in the southern
lobe. The reasons for this treatment are discussed in L97.
\end{itemize}

\subsubsection{Hot spots}

As in H97, we attempt to identify a `primary' hot spot in each
lobe. Flux and size measurements are tabulated for the primary
components only. Cases where the identification of the primary hot
spot might be ambiguous are discussed in the notes to Table \ref{bhot_tab}.

B94 measure the flux densities and dimensions of the hot spots in
their quasar sample by fitting with a Gaussian and baseline. They
remark that such models do not always represent the hot spots well,
and the situation is still more difficult in the present sample with
its considerably higher spatial resolution and with a more relaxed
definition of the term `hot spot'. The procedure we have adopted is as
follows, therefore: we have tried to fit models consisting of a
Gaussian and baseline to all but the most obviously resolved hot
spots, running the fitting routine a number of different times on
slightly different regions or with slightly different initial
guesses. Where these different fits gave essentially the same results,
we have tabulated in Table \ref{bhot_tab} the average values of the
integrated flux densities and major and minor axis lengths; the error
on the flux densities is estimated from the range of the results of
different measurements. Where the fits were obviously poor, and
diverged significantly (by a factor of 1.5 or more in flux density
either way) on small alterations of the initial parameters of the
fitting routine, we measured the sizes of the components by taking
slices through them and estimating the FWHM, or in extremely resolved
cases by simple measurement from maps; we measured the flux densities
by integration from the maps, with a rough correction for background
emission from integration over a nearby part of the lobe. The errors in
the derived hot spot flux densities in these cases are an estimate of
the error from direct integration, taking into account the difficulty
of deciding exactly which was the hot spot region.

The distances of the hot spots from the cores are also tabulated.

\section{Trends}
\label{trends}

The high-frequency flux densities we use in this section have all been
corrected to an observing frequency of 8.4 GHz for ease of
comparison. Cores are assumed to have flat spectra ($\alpha=0$) and
jets to have $\alpha = 0.8$. Total flux densities are scaled using the
low-frequency (178-750 MHz) spectral index, which is available
for the whole sample. Errors in these corrections should not badly
affect the results of the analysis performed here.

Uniform symbols are used in the plots in this section. Open circles
represent LERG, open triangles indicate NLRG, filled
triangles indicate BLRG and filled stars indicate quasars. The one
unclassified object, \Ss{3C136.1}, is plotted as a dotted circle and
is not included with any class of object where the objects are
separated by emission-line type.

The `prominence' of a feature is defined throughout as the ratio of
the flux density of a given feature to the extended flux of the source
(i.e. the total flux density of the source minus contributions from
jets and cores). This is a useful quantity for indicating effects due
to relativistic beaming, since the extended flux density of the source
should be orientation-independent.

The median values of some of the distributions discussed below are
tabulated, broken down by object class, in Table \ref{median}. For
comparison, the median values using only objects drawn from the
complete sample of LRL are also tabulated. It will be seen that in
general the addition of the non-LRL objects makes little difference to
the distributions. All the results that follow are based on the full sample.

\subsection{Large-scale structure}

\begin{figure*}
\begin{center}
\leavevmode
\epsfysize=9cm\epsfbox{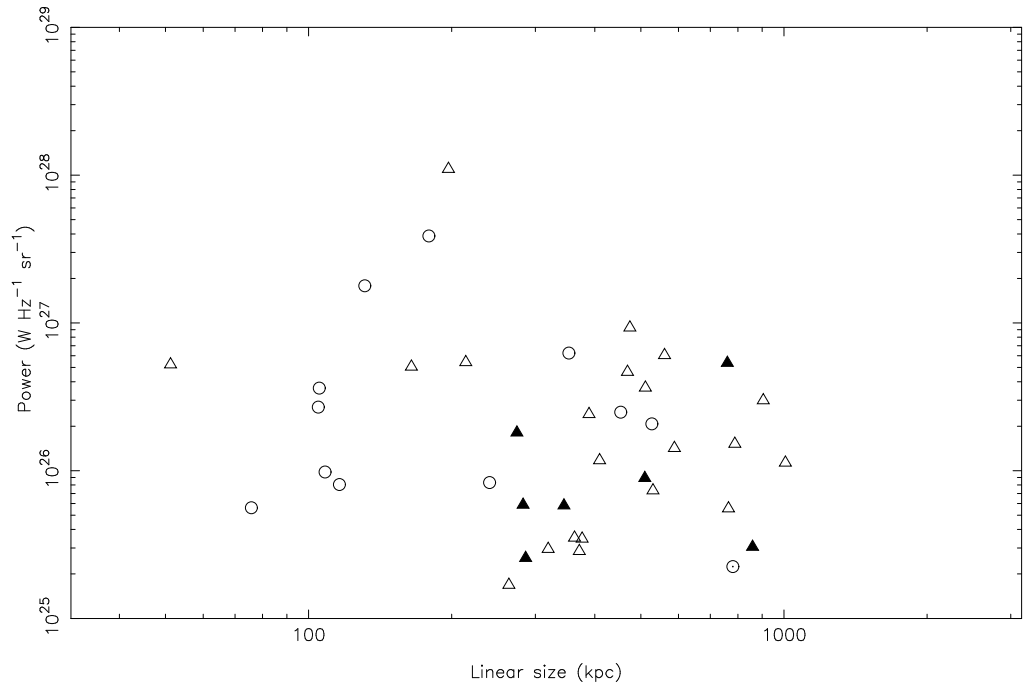}
\caption{The power-linear-size diagram for the sample of radio
sources. Open circles denote LERG; a dotted circle denotes an
unclassified object; open triangles denote NLRG and filled triangles BLRG.}
\label{pdd}
\end{center}
\end{figure*}

Fig.~\ref{pdd} shows the power-linear-size diagram for the whole
sample. The sources populate a range of two decades in luminosity, if
the brightest objects (Cygnus A\Ssq{3C405}, \Ss{3C123} and
\Ss{3C438}) are not counted, and approximately 1.5 decades in linear
size (the upper limit in this plot is due to the exclusion of the
giants; the lower limit is due to the absence of compact sources from
LRL at low redshifts).

It will be seen that there is a tendency for the LERG to be smaller;
the distribution of linear sizes of these objects is strongly peaked
around the 100-kpc range, with a median of 130 kpc compared to 370 kpc
in the sample as a whole. A Wilcoxon-Mann-Whitney test rejects the
hypothesis that the LERG in this sample are drawn from a size
distribution with the same median as that of the other objects at the
99.9 per cent level; even if the giant radio galaxies (some of which
are LERG) are included in the test, the samples are still
significantly different. We shall return to this initially surprising
fact, first noted by Black (1992) for his sub-sample of these objects,
in a further paper. The median linear size of BLRG objects is smaller
than that of NLRG but the difference between the two distributions is
not significant.  In the larger and higher-power sample of L94 the
broad-line objects are significantly smaller.

The median linear size of the B94 quasars is comparable to that of
the radio galaxies (Table \ref{median}). However, the B94 sample is biased
towards larger linear sizes because of the angular size selection
criterion, so this comparison is probably not very meaningful.

\subsection{Hot spots}

\begin{figure*}
\begin{center}
\leavevmode
\epsfysize=12cm\epsfbox{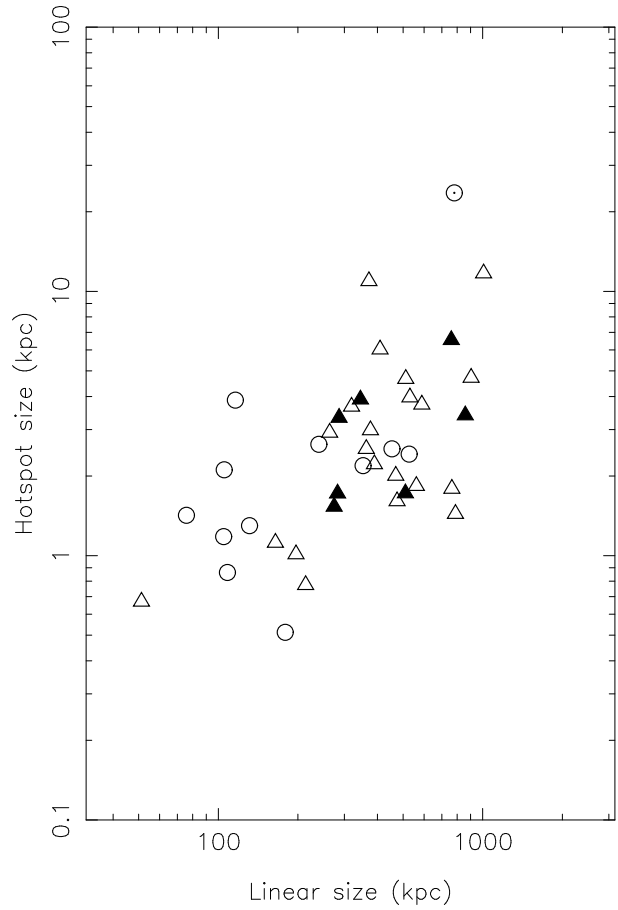}
\caption{Linear sizes of the hot spots of the objects with good
measurements against largest linear size. Open circles denote LERG; a
dotted circle denotes an unclassified object; open triangles denote
NLRG and filled triangles BLRG.}
\label{lshs}
\end{center}
\end{figure*}

There is a strong positive correlation (significant at the 99.9 per
cent level on a Spearman Rank test), with slope approximately unity,
between linear size and hot spot size in the 43 sources in the sample
with measured hot spots (Fig.\ \ref{lshs}). Hot spot size is defined
here as the average of the sizes of the two hot spots at either end;
the size of an individual hot spot is defined as the geometric mean of
the largest and smallest angular sizes. Care must be taken when
comparing the sizes of hot spots in different objects, since they have
been measured from maps of differing angular resolution; objects of
large angular size are more likely to have been mapped at lower
resolution in this sample. However, the correlation is also present
when only those measurements made from maps of $\sim 0.23$ arcsec
resolution are considered, suggesting that an intrinsic correlation
exists. This result provides some support for suggestions in Laing
(1989) and B94 that the hot spot sizes scale with source linear size;
however, Black (1992) found no such correlation in the sub-sample of
these objects that he analysed. If the correlation is real, it is
consistent with self-similar models for these sources (e.g.\ Kaiser \&
Alexander 1997) or may be evidence for `tired jet' models in which the
jet decelerates with distance from the core (B94). It is evidence
against models in which the beam size is independent of the linear
size of the source (e.g.\ magnetic self-confinement).

Comparing the sizes of hot spots within a single source does not
suffer from these difficulties. Of the 32 sources with detected jets
or possible jets and measurements of hot spot size, 24 had hot spots
that differed in area by more than 25 per cent. Of these 15 (62 per
cent) had the brighter or only jet pointing towards the more compact
hot spot. This lack of a significant trend contrasts with the results
of B94 who found that, where the hot spots were significantly
different in size, the jets in their sample of quasars always pointed
towards the more compact hot spots. They found a dependence of hot
spot size ratio on core power which is absent in this sample. B94 also
found a trend for the hot spot on the jetted side to be more recessed
(where recession is measured by the ratio of the core-hot spot
distance to the lobe length) whereas in the present sample the jetted
hot spot was {\it less} recessed in 21 (65 per cent) of the 32
sources, a marginally significant trend (significant at the 90 per
cent level on a binomial test).

22 sources with jets in the present sample had one hot spot brighter
than the other within the errors assigned. The brighter or only jet
pointed towards the dimmer hot spot in 7 cases and towards the
brighter in 15 (68 per cent) -- a marginally significant trend
(significant at the 90 per cent level on a binomial test). This may be
contrasted with the results of Laing (1989), who found that 26/30
sources with jets had the jet pointing towards the brighter hot spot
in a sample of powerful radio galaxies and quasars. It would appear
that all the results relating to hot spots are weaker, if present at
all, in this sample than in high-redshift samples containing quasars.
This is understandable if the trends in the B94 sample are due to
relativistic effects (as suggested by theory; e.g.\ Komissarov \&
Falle 1996), since at least some of the objects studied here cannot be
strongly beamed towards us.

\begin{figure*}
\begin{center}
\leavevmode
\epsfxsize=\linewidth\epsfbox{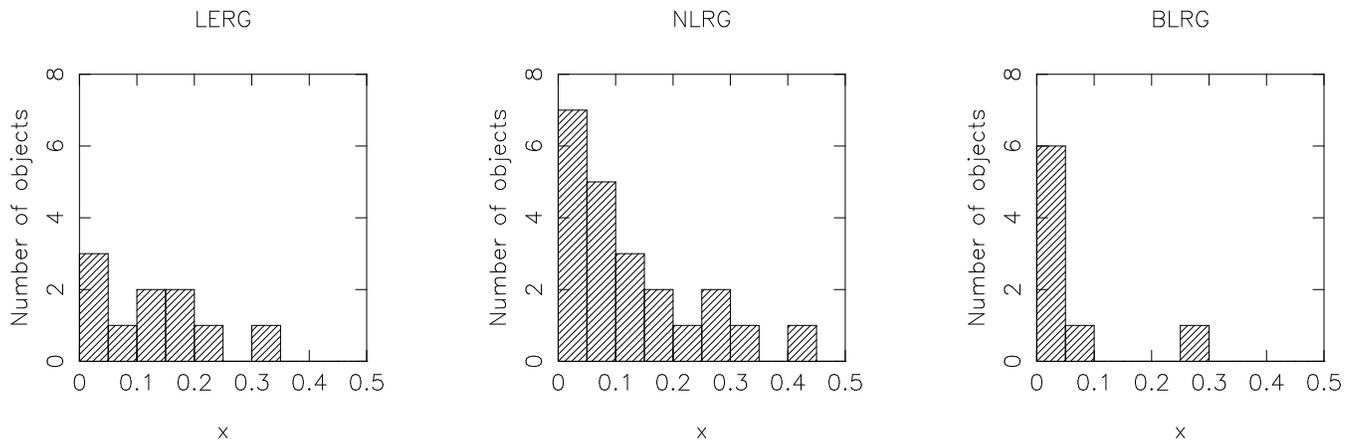}
\caption{Histograms of the fractional separation parameter $x$ for the
sources with measured hot spot positions. The unclassified source is
not plotted.}
\label{asyp}
\end{center}
\end{figure*}

Best \etal\ (1995) have analysed the hot spot separation ratios of a
large sample of FRII radio sources (with some overlap with the present
sample, but including many high-power sources), showing that there are
significant differences between the distributions of radio galaxies
and quasars. The quasars are more asymmetrical, a fact that they
attribute to mildly relativistic hot spot advance speeds. They
concluded that in the low-redshift r\'egime there was no difference
between the hot spot separation ratios of BLRG and NLRG, but were
unable to test whether the LERG formed a distinct population. In Fig.\
\ref{asyp} we show histograms of the fractional separation difference
$x$ [$x = (\theta_1 -\theta_2) / (\theta_1 + \theta_2)$, where
$\theta_1$ and $\theta_2$ are the lengths of the longer and shorter
lobe respectively] for the 40 objects in this sample with measured hot
spot positions. It will be seen that the distributions are not
markedly different for the NLRG and the LERG, and a
Wilcoxon-Mann-Whitney test finds no significant probability that the
two are drawn from different distributions. The broad-line objects
appear {\it more} symmetrical than the narrow-line objects (a result
significant at the 97 per cent level on a Wilcoxon-Mann-Whitney
test). These results are both rather surprising; in the simple models
outlined above we would expect BLRG to be more asymmetrical than the
NLRG, because of their smaller mean angle to the line of sight, while
LERG should also be more asymmetrical than NLRG if they are an
isotropically distributed population. This may be evidence that in
this comparatively low-redshift and low-power r\'egime environmental
effects on source symmetry are dominant over light travel time
effects.

\subsection{Jets}
\label{discuss-jets}

\begin{figure*}
\begin{center}
\leavevmode
\epsfysize=9cm\epsfbox{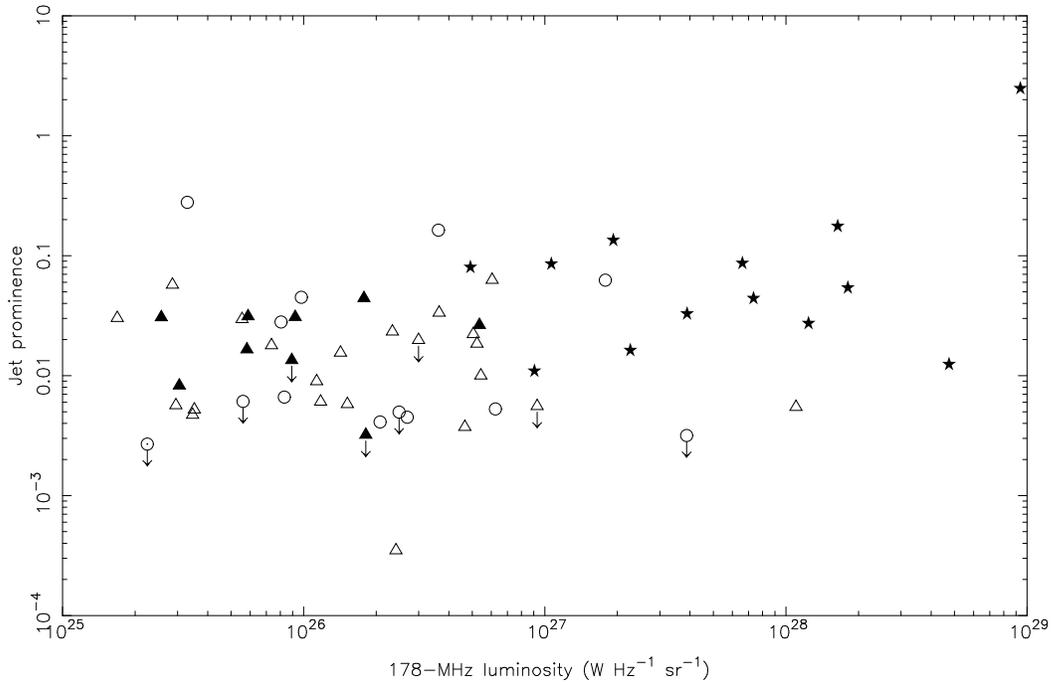}
\caption{(Brighter) jet prominence against total low-frequency luminosity. Upper
limits are marked with arrows. The radio galaxies are shown with the
B94 quasars. Open circles denote LERG; a
dotted circle denotes an unclassified object; open triangles denote
NLRG and filled triangles BLRG; filled stars denote quasars.}
\label{jetl}
\end{center}
\end{figure*}

In the sample as a whole 29/50 (58 per cent) of objects have definite
jet detections and 40/50 (80 per cent) have possible or definite jets.
The jet detection rates are not significantly different in the two
sub-samples of B92 and H97.  There appears to be no dependence of jet
detectability on luminosity over the luminosity range studied here;
this is shown in Fig.\ \ref{jetl}, a plot of jet prominence against
luminosity for the 44 sources with measured jets or possible jets or
upper limits together with the B94 quasars.  In all but two cases
radio galaxy jets have less than 6 per cent of the extended flux
density of the source, and the typical value is nearer to 1 per cent.
Parma \etal\ (1987) found an inverse correlation between jet
prominence and luminosity in a sample of lower-power (mostly FRI)
objects, but we find no such strong relationship in FRIIs over our
luminosity range. Where the luminosity ranges overlap our results are
consistent with those of Parma \etal , so it seems likely that their
results are indicative of an increase in jet efficiency across the
FRI/FRII boundary which has levelled out by the luminosity range of
our sample.

Jets are one-sided. In the sample as a whole 16 objects (32 per cent)
have possible counterjet detections, but only 6 (12 per cent) have
definite detections, these being the unusual objects \Ss{3C15},
\Ss{3C171} and \Ss{3C438}, the bright, deeply imaged
objects \Ss{3C405} and \Ss{3C353}, and the highly symmetrical source
3C\,452. The counterjet candidate detection fraction is thus below
that of B94 in their sample of quasars (54 per cent have counterjet
candidates), which is a surprising result in the context of unified
models.

In the only independent search for jets recently conducted, Fernini
\etal\ (1993) imaged a sub-sample of five objects with powers matched
to those of the B94 quasars, and found only one definite jet (but
three possible jets). Although the sensitivity in this study should
have been comparable to that of our observations, the spatial
resolution was lower by a factor of 2 because of the higher redshift of
these objects and the lower angular resolution of the
observations. Too much should not be read into the differing jet
detection fractions, as the sample is small.

Perhaps surprisingly, the jet detection fraction is not very different
in the different classes of object studied here. Of the 15
low-excitation objects, 9 (56 per cent) have definite jets and 11 (73
per cent) have possible or definite jets. For the 9 BLRG, the figures
are 7 (78 per cent) with definite jets and the same number with
definite or possible. The 25 NLRG have 13 definite jets (52 per cent)
and 22 definite or possible (88 per cent). None of these jet detection
fractions are sufficiently different to provide evidence that the
different classes of object are drawn from different populations.  L94
found a much more marked difference between the jet detection
fractions of broad-line objects (including quasars) and NLRG in their
higher-power sample.

It is perhaps unexpected, in the model in which the BLRG are quasar
counterparts, that there should be two (\Ss{3C227} and
\Ss{3C381}\footnote{The classification of 3C\,381 as a BLRG is
uncertain; see H97.}) with no indication of a jet -- compare the
universal jet detection in the B94 quasars. On the other hand, several
of the BLRG that do have jets are very similar in appearance to the
quasars. The median straight jet prominence of the BLRG is twice that
of the NLRG (Table \ref{median}) and the difference between the
samples is significant at the 95 per cent level on a median test (the
only simple test that can be applied rigorously, given the number of
upper limits in the data).

Fig.\ \ref{jetl} shows that the distribution of BLRG total jet
prominence overlaps with the low end of quasar total jet prominence;
the B94 quasars include objects with total jet prominence much higher
than in either the BLRG or the NLRG, and this is reflected in the
higher median total jet prominence of the quasars (Table
\ref{median}); the difference between BLRG and quasars is significant
at the 95 per cent level on a median test. However, the prominences of
{\it straight} jets in the BLRG and the B94 quasars are similar; the
distributions cannot be distinguished on a median test.

\begin{figure*}
\begin{center}
\leavevmode
\epsfysize=9cm\epsfbox{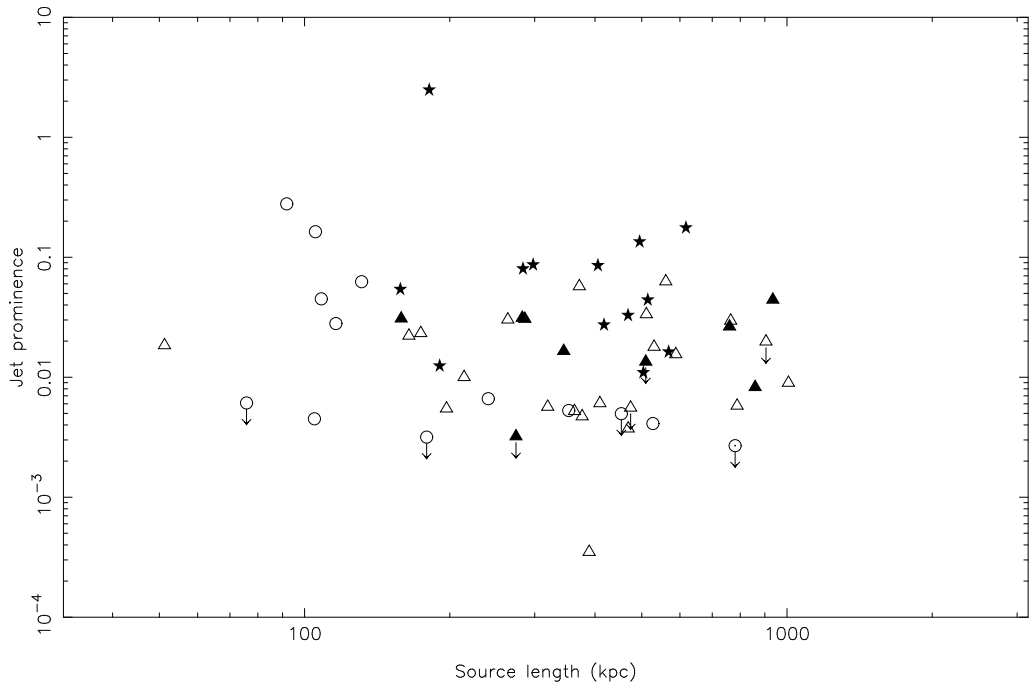}
\caption{(Brighter) jet prominence against linear size. Upper
limits are marked with arrows. The radio galaxies are shown with the
B94 quasars. Open circles denote LERG; a dotted
circle denotes an unclassified object; open triangles denote NLRG and
filled triangles BLRG.}
\label{jetlen}
\end{center}
\end{figure*}

Black (1992) noted a tendency for jets to be detected in shorter
objects. In the present sample 24 of the 40 radio galaxies with
definite or possible jets (60 per cent), have linear sizes less than
the median size for the whole sample of 370 kpc, a trend significant
at the 90 per cent level. These results are weaker than those of Black
(1992), probably because of the detection or possible detection of a
number of faint jets in larger sources since his analysis. Fig.\
\ref{jetlen} shows that the weak trend for jet prominence to be
related to length is entirely because of the existence of a number of
short, low-excitation objects with prominent jets; 3C\,15 and 3C\,401
are examples of this class, which overlaps in its radio properties
with the `jetted double' class of Law-Green \etal\ (1995), as
discussed by H97.

There is no significant tendency for the
brighter or only jet to lie in the longer lobe (18/33) or the brighter
lobe (16/26). Such trends might be expected if relativistic and
light-travel-time effects dominated the appearance of radio sources.

The lengths of the jets and straight jets are strongly correlated
($>99$ per cent significant on a Spearman Rank test) with the total
length of the source. This is also consistent with self-similar models
for the sources, and militates against models which define a scale
length for the jet or the onset of jet bending (e.g.\ any model which
relates these quantities to galactic scale sizes). There is no
significant correlation between straight jet flux density and length,
however.

\subsection{Jet sidedness}
\label{sided_m}
The jet-counterjet ratio should be an indicator of the velocities
involved and the projection angles in models based on relativistic
beaming. If the jet and counterjet are intrinsically symmetrical, then
their relative flux densities are found from equation (\ref{db}). For
the reasons given in section \ref{strjet}, the straight segments of
jets are best for this kind of test. In general the ratio $J$, the
straight jet sidedness, is given by

\begin{equation}
J = R \left({{1 + \beta_j \cos \theta}\over{1-\beta_j \cos
\theta}}\right)^{2 + \alpha_j}
\label{jcj}
\end{equation}
where $R$ is the degree of {\it intrinsic} asymmetry (the ratio
between the rest brightnesses of jet and counterjet) and might
represent the effects of different environments on the efficiencies of
the two beams (neglecting the effects such environments might have on
velocities). It follows from equation (\ref{jcj}) that, if relativistic
beaming effects are dominant in the jets, objects at smaller angles to
the line of sight should have larger jet-counterjet asymmetries.

\begin{figure*}
\begin{center}
\leavevmode
\epsfysize=18cm\epsfbox{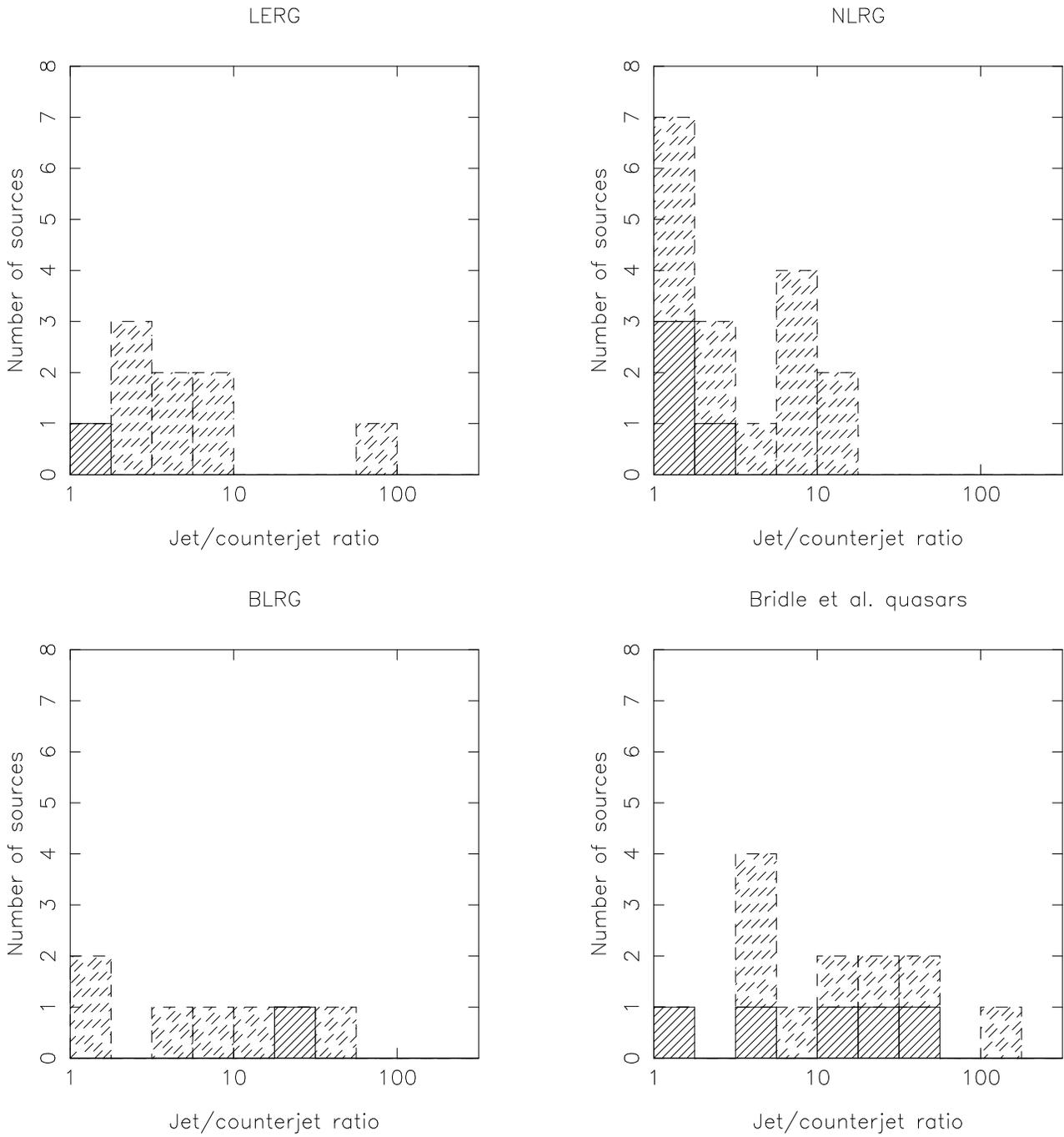}
\caption{Jet sidedness
of radio galaxies and quasars. Dashed shading indicates lower
limits.}
\label{sided}
\end{center}
\end{figure*}

Fig.\ \ref{sided} shows the distribution of straight jet sidednesses
for the radio galaxies with detected jets and for the B94
quasars. Dashed areas of the histograms represent {\it lower} limits on
the jet-counterjet ratios (since they are calculated using upper limits
on the counterjet flux density). Clearly selecting on detected jets
biases the sample; in the context of unified models, sources without
detected jets will be preferentially unbeamed and so the more
two-sided sources are excluded from the plot.

The plot as it stands can at best be said to be consistent with
unified models (which would not be the case if, for example, the
radio galaxies were obviously more one-sided than the quasars).
However, the paucity of counterjet detections in the radio galaxy
population is surprising if unified models are correct (though the
counterjet detection fraction is underestimated by this plot which
excludes non-aligned counterjets in, for example, \Ss{3C15} and \Ss{3C438}).

\subsection{Cores}
\label{cores}

The prominence of the radio core is commonly used as an orientation
indicator in studies of beaming in radio galaxies and quasars (e.g.\
Orr \& Browne 1982). Various authors
(e.g.\ Kapahi \& Murphy 1990; L94; Morganti \etal\ 1995; Morganti
\etal\ 1997) have attempted to show consistency between the
distribution of core prominences and the predictions of unified
models.

In the present sample, the median core prominence of BLRG is nearly an
order of magnitude greater than that of the NLRG (Table \ref{median})
and a Wilcoxon-Mann-Whitney test allows us to reject the hypothesis
that the core prominences for the two classes are drawn from a
distribution with the same median at $>99$ per cent
confidence. Morganti \etal\ (1997) find a similar difference between
the median core prominences of BLRG and NLRG.  The core prominences
for both broad-line and narrow-line objects have a scatter of 2--2.5
orders of magnitude.

The LERG have a median intermediate between the broad- and narrow-line
objects (as they do in L94 and in the largely disjoint sample of
Morganti \etal\ 1997) and a Wilcoxon-Mann-Whitney test does not
distinguish the median core prominence of the LERG from that of the
high-excitation objects with significant probability. In the simple
versions of unified models discussed above (section \ref{unif}) we
might expect the core prominences of LERG to be distributed like those
of the broad- and narrow-line objects combined, which does not seem to
be the case in this sample (there is a lack of low-excitation objects
with very bright cores). This result is only marginally significant
(at the 90 per cent level on a Kolmogorov-Smirnov test).

The core prominences of BLRG and B94 quasars are similar;
the medians are very close (Table \ref{median}) and the distributions
are not distinguished by a Wilcoxon-Mann-Whitney test. This result
differs from that of Morganti \etal\ (1997) who found a significant
difference between the core prominences of FRII quasars and BLRG.
We discuss this further in section \ref{disc-beaming}.

\begin{figure*}
\begin{center}
\leavevmode
\vbox{\epsfysize=9cm\epsfbox{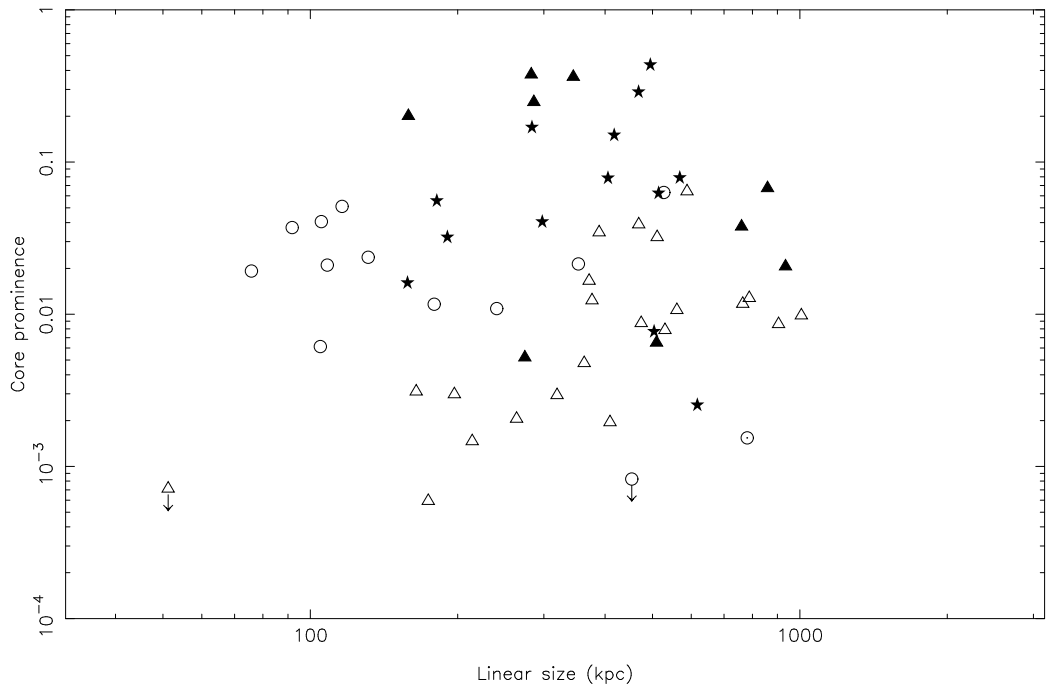}}
\caption{Core prominence as a function of source linear size. The radio
galaxies are shown with the B94 quasars. Open circles denote LERG; a
dotted circle denotes an unclassified object; open triangles denote
NLRG and filled triangles BLRG; filled stars denote quasars.}
\label{cplen}
\end{center}
\end{figure*}

There is no apparent trend in the core prominences as a function of
luminosity. However, on a plot of core prominence against linear size
(Fig.\ \ref{cplen}) the NLRG (only) show an apparent positive trend
(significant at $>99$ per cent on a Spearman Rank test); i.e. for
NLRG, a longer source tends to have a more prominent core. There is no
tendency for the extended flux densities in these objects to
anticorrelate with length, so this trend, though surprising, must be
real.

\begin{figure*}
\begin{center}
\leavevmode
\vbox{\epsfxsize=10cm\epsfbox{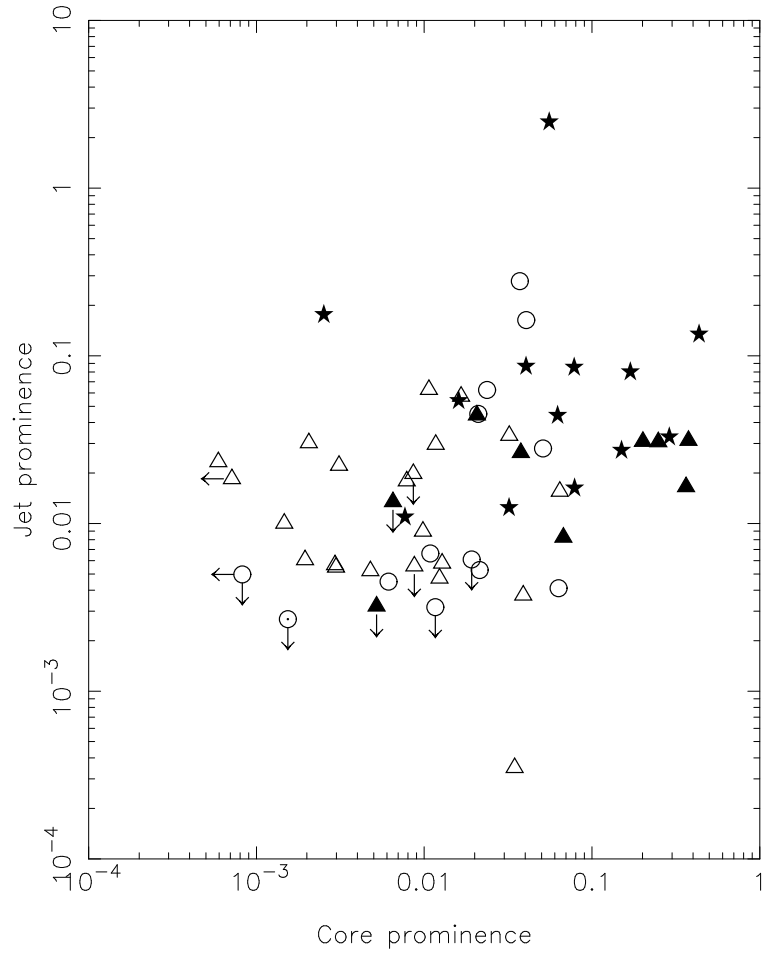}}
\caption{(Brighter) jet prominence as a function of core
prominence. The radio galaxies are shown with the B94
quasars. Open circles denote LERG; a
dotted circle denotes an unclassified object; open triangles denote
NLRG and filled triangles BLRG; filled stars denote quasars.}
\label{tjco}
\end{center}
\end{figure*}

If we believe that the core is a beamed parsec-scale jet (as suggested
by VLBI observations) and that the one-sidedness of jets is a
relativistic effect, there should in principle [equation (\ref{db})]
be a relationship between the prominence of jet and core, assuming
that the beam does not change direction significantly between parsec
and kiloparsec scales.  In general, the velocity in the core may not
be equal to that in the kiloparsec-scale jet, and so the slope of a
jet-core prominence relation will not be unity. B94 investigate the
relation between core and jet prominence for their sample of quasars,
and conclude that there is such a relationship, with deceleration
required between the parsec- and kiloparsec- scales. Fig.\ \ref{tjco}
shows a plot of total jet prominence against core prominence for the
present sample.

B94 chose to investigate the relationship between the core and {\it
straight} jet prominence. They point out that a well-defined
relationship between the core and jet beaming factors only occurs when
the two are at the same angle to the line of sight. As soon as the jet
is seen to bend, the emission is beamed differently and the simple
relationship is lost. This could be used to explain the absence of a
strong relationship in Fig.\ \ref{tjco}.

\begin{figure*}
\begin{center}
\leavevmode
\vbox{\epsfxsize=10cm\epsfbox{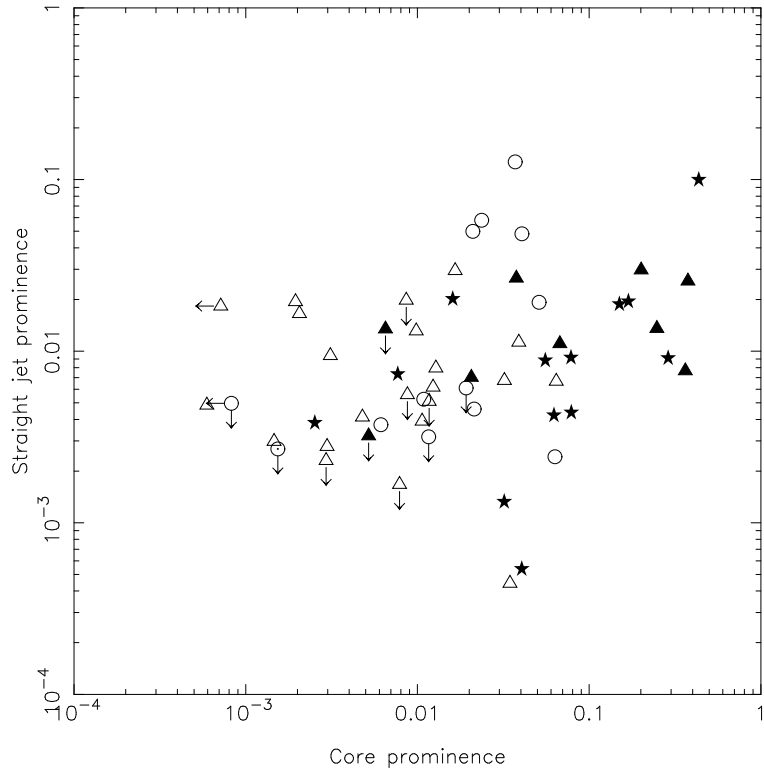}}
\caption{(Brighter) straight jet prominence against core
prominence for the radio galaxies and B94 quasars. Upper limits are marked
with arrows. Open circles denote LERG; a dotted circle denotes an
unclassified object; open triangles denote NLRG and filled triangles
BLRG; filled stars denote quasars.}
\label{stjco}
\end{center}
\end{figure*}

In Fig.\ \ref{stjco} we plot the relationships between (brighter)
straight jet and core prominences for the well-mapped sources in the
present sample and for the quasars of B94. Unlike them, we have made
no attempt to assign some of the flux density from the core to the
straight jet, as we have little information on the VLBI properties of
the radio galaxies. This weakens the correlation observed by B94 for
their sample, though it is still significant at the 90 per cent level
on a Spearman Rank test. Because the cores and straight jets of the
BLRG and of the quasars are more prominent than those of the NLRG, as
discussed above, there is a weak but significant positive trend in
Fig.\ \ref{stjco}, even in the presence of upper limits [$>95$ per
cent on a generalised Spearman Rank test, as implemented in the survival
analysis software package {\sc asurv} Rev.~1.1 (LaValley, Isobe \&
Feigelson 1992)]. This is true whether or not the B94 quasars are
included in the correlation analysis.

Also noticeable in Fig.\ \ref{stjco} is a population of LERG whose
straight jets are much brighter than those of other objects of similar
core prominence.

\section{Discussion}

We summarise the important results of section \ref{trends} in Table
\ref{summary}. In this section of the paper, we discuss some
interesting general properties of the sources that have emerged from
our analysis, examine the evidence for unification in this sample, and
consider ways in which the data can be modelled.

\subsection{General radio galaxy properties}

Jets or possible jets are detected in most of the sources in the
sample, and we have been able to measure, or put upper limits on, the
background-subtracted flux densities of the jets of the objects in the
sample for which we have good maps.  There is no strong dependence of
the prominences of these jets either on the luminosity of the radio
galaxy or on its linear size. These results make it seem likely that
with even better observations we should be able to detect jets in all
FRII radio galaxies.  Further observations of higher-luminosity radio
galaxies, which are more directly comparable to quasars, are necessary
to investigate these results further.

Core prominence shows no trend with luminosity in the present sample
(but cf.\ the anticorrelation seen in the larger samples, spanning a
wider luminosity range, of Giovannini \etal\ 1988 and Zirbel \& Baum
1995). In the NLRG alone, core prominence is positively correlated
with source length (the BLRG and quasars and the LERG show no
significant correlation). Source length is affected by angle to the
line of sight, but the dominant contribution to the distribution of
source lengths must be intrinsic (in any case, a simple beaming model
with constant-length sources would predict an {\it anticorrelation}
between source linear size and core prominence). The normalising
extended flux density is not correlated with linear size. It appears therefore
that there is a direct relation between core prominence and total
source length, at least in the NLRG; relativistic beaming and
projection effects might be expected to wash out such an effect in the
BLRG and quasars. The physics behind such a correlation is unclear,
but it will be interesting to see whether it persists in larger samples.

Both the linear sizes of the hot spots and the lengths of the straight
jets are strongly, and approximately linearly, correlated with the total
linear sizes of the radio source. This seems to rule out models which
set scale sizes on either the width of the beam or the distance it can
travel without disrupting.

\subsection{Unified models and relativistic velocities}
\label{disc-beaming}

In section \ref{unif} we pointed out that the absence of
lobe-dominated quasars from our sample {\it requires} that some of
these radio galaxies be oriented at small angles to the line of
sight. We outlined a model in which some or all of the BLRG in the
sample are the low-redshift counterparts of quasars.

We have seen that the BLRG have systematically brighter cores and
straight jets than the NLRG, and that the distributions of these
quantities and of jet sidedness in the BLRG are very similar to those
of the B94 quasars (although, as pointed out above, those objects are
not an ideal population for the comparison). These results are
consistent with a model in which the cores and jets in BLRG are
preferentially beamed [as expected in most unified models and
concluded, in the case of cores, by Morganti \etal\ (1997)] and in
which the BLRG are similar in their orientations to quasars.

There remain a number of problems with simply identifying all
broad-line radio galaxies as low-power lobe-dominated quasars. The
facts that some BLRG in our sample have no detected jets, and that the
BLRG appear more symmetrical than the NLRG, may perhaps be attributed
to random environmental effects, given the size of the sample, or to
misclassification. The fact that the {\it total} jet prominences of
the B94 quasars are very much higher than those of our BLRG is
interesting; there appear to be no BLRG in our sample comparable to
the jet-dominated objects in the B94 sample (e.g.\ 3C\,9). Since these
results imply that the quasar jets are significantly brighter after
they bend, it is hard to explain it as a simple relativistic beaming
effect; it may indicate real differences in jet efficiency, power or
environment between the low-redshift BLRG and the high-redshift
quasars.

As discussed above, Morganti \etal\ (1997) found that BLRG cores in
their sample were systematically less prominent than those of
lobe-dominated quasars, even after selection effects were taken into
account. However, as they point out, the redshift distributions for
the BLRG and quasars in their sample are very different, and their
sample, like ours, contains, at $z\la 0.3$, very few lobe-dominated quasars
and many BLRG. At high redshifts BLRG and quasars are known to
coexist, and our results are not inconsistent with the suggestions of
Barthel (1989) and Morganti \etal\ (1997) that high-$z$ BLRG are
transitional objects.

When the prominences of straight jets and cores are compared (Fig.\
\ref{stjco}), a weak positive trend can be seen in the sense that
objects with more prominent cores tend to have more prominent
jets. This trend is unlikely to be much affected by the fact that some
of the points are upper limits. The trend is in agreement with the
results of B94, and the slope of the correlation is certainly less
than unity. Such a result is expected if, as suggested by B94, both
the cores and jets are beamed, with the emitting material in the jets
having an effective bulk velocity lower than that seen in the
core. Given that we expect the cores to be beamed for other reasons,
it is hard to explain such a correlation without relativistic
velocities in the radio galaxy jets as well. The sidednesses of the
jets, when compared to those of the jets in B94, are also consistent
with such a model, although the result is weak since there are few
good counterjet detections.

Our results on the core prominence of LERGs are broadly consistent with the
suggestion that they form an isotropic population, as found by L94. 
However, there is (in this sample) a paucity of LERG with
very bright cores, and there is also a sub-class of these objects with
very prominent jets and diffuse, weak hot spots; the LERG are also
significantly smaller than the high-excitation objects. These results
suggest that there is some physical difference, either intrinsic or
environmental, between the radio properties of the LERG and the other
objects; we shall explore the reasons for this difference elsewhere.

We find no relationship between the jet side and the more compact or
the brighter hot spot; this is consistent with the suggestion that
these effects, seen by B94 and L89, are due to relativistic beaming in
the hot spot and that the present sample is less beamed than those
previously investigated because of a wider range of angles to the
line of sight.

\subsection{Modelling: future work}

These results allow us to fit empirical models to the jet and core
prominence data and the jet sidedness data, with the input being the
distribution of angles to the line of sight and the intrinsic length
distribution of sources; the ability to ignore the luminosity
distribution and its relation to linear size is an important
simplification. Free parameters to a model fit would include the bulk
velocities in jets and cores, a parametrisation of the dependences of
the prominences of both on linear size, and some normalising
`intrinsic prominence', corresponding to the unbeamed prominence of
the jet and core; in addition, some parametrisation of the
distribution both of velocities and of intrinsic prominences (which
might depend on source environment) might be necessary. Such model
fitting has been done before on projections of datasets of this type
(e.g.\ Morganti \etal\ 1995; Wardle \& Aaron 1997) but we are in a
position to use the relationships between the measured quantities, and
so to put tighter constraints on the allowed regions of parameter
space. We shall discuss results of this type of analysis in a further
paper.

\section*{ACKNOWLEDGEMENTS}

MJH acknowledges a research studentship from the UK Particle Physics
and Astronomy Research Council (PPARC) and support from PPARC grant
GR/K98582. We are grateful to Alan Bridle, Jack Burns, David Clarke,
Jane Dennett-Thorpe, Paddy Leahy, Mark Swain and Wil van Breugel for
supplying images used in the analysis, and thank Peter Scheuer and
Christian Kaiser for useful discussions. We thank an anonymous referee
for a number of helpful suggestions which allowed us to improve the
clarity of the paper.

The National Radio Astronomy
Observatory is operated by Associated Universities Inc., under
co-operative agreement with the National Science Foundation. This
project made use of STARLINK facilities. This research has made use of
the NASA/IPAC Extragalactic Database (NED) which is operated by the
Jet Propulsion Laboratory, California Institute of Technology, under
contract with the National Aeronautics and Space Administration.

\bsp

\begin{table*}
\caption{The combined sample of FRII radio sources}
\label{combsamp}
\begin{center}
\begin{tabular}{lld{4}rrrd{1}rllll}
Source&IAU name&z&$S_{178}$&$\alpha$&$P_{178}$&\multicolumn{1}{r}{LAS}&Size&Jets?&Ref.&Freq.&Emission\\
&&&(Jy)&&&\multicolumn{1}{r}{(arcsec)}&(kpc)&&&(GHz)&class\\
\hline
\bf\Ss{4C12.03} &0007+124& 0.156  &  10.9  & 0.87  &  104  &215& 787 &$\circ\circ$&(9)&--&E \\
\Ss{3C15} &0034$-$014&0.0730  &  17.2  & 0.67  &   33  & 48.0&  92 &$\bullet\bullet$&2&8.4&E\\
\bf\Ss{3C20} &0040+517& 0.174  &  46.8  & 0.66  &  543  & 53.6& 214 &$\bullet\circ$&1&8.4&N \\
\bf\Ss{3C33} &0106+130&0.0595  &  59.3  & 0.76  &   75  &254& 404  &&(9)&--&N\\
\bf\Ss{3C33.1} &0106+729& 0.181  &  14.2  & 0.62  &  178  &227& 935 &$\bullet$&11&1.5&B \\
\bf\Ss{3C61.1} &0210+860& 0.186  &  34.0  & 0.77  &  462  &186& 782&$\circ$&(12)&--&N  \\
\bf\Ss{3C79} &0307+169&0.2559  &  33.2  & 0.92  &  930  & 89.0& 474 &&1&8.4&N \\
\bf\Ss{3C98} &0356+109&0.0306  &  51.4  & 0.78  &   17  &310& 264 &$\bullet$&2&8.4&N \\
\Ss{3C105} &0404+035& 0.089  &  19.4  & 0.61  &   56  &335& 764 &$\circ$&2&8.4&N  \\
\bf\Ss{4C14.11} &0411+141& 0.206  &  12.1  & 0.84  &  208  &116& 527&$\bullet$&1&8.4&E  \\
\Ss{3C111} &0415+379&0.0485  &  70.4  & 0.76  &   59  &215& 283 &$\bullet$&2&8.4&B \\
\bf\Ss{3C123} &0433+295&0.2177  & 206.0  & 0.70  & 3873  & 37.8& 179&&1&8.4&E  \\
\bf\Ss{3C132} &0453+227& 0.214  &  14.9  & 0.68  &  269  & 22.4& 105&$\circ$&1&8.4&E  \\
\bf\Ss{3C135} &0511+008&0.1273  &  18.9  & 0.95  &  118  &132& 409 &$\bullet$&2&8.4&N \\
\Ss{3C136.1} &0512+248& 0.064  &  15.3  & 0.72  &   22  &460& 781 &&2&8.4&? \\
\bf\Ss{3C153} &0605+480&0.2771  &  16.7  & 0.66  &  524  &  9.1  &  51 &$\circ\circ$&1&8.4&N \\
\bf\Ss{3C171} &0651+542&0.2384  &  21.3  & 0.87  &  505  & 32.5& 165 &$\bullet\bullet$&1&8.1&N \\
\bf\Ss{3C173.1} &0702+749& 0.292  &  16.8  & 0.88  &  624  & 60.5& 353&$\bullet$&1&8.4&E  \\
\bf\Ss{3C184.1} &0734+805&0.1182  &  14.2  & 0.68  &   73  &182& 530 &$\circ$&2&8.4&N \\
\bf\Ss{3C192} &0802+243&0.0598  &  23.0  & 0.79  &   29  &200& 319 &$\circ$&2&8.4&N \\
\Ss{3C197.1} &0818+472&0.1301  &   8.8  & 0.72  &   56  & 24.0&  76 &&3&8.4&E \\
\bf\Ss{3C219} &0917+458&0.1744  &  44.9  & 0.81  &  536  &190& 760 &$\bullet\circ$&10&4.9&B \\
\bf\Ss{3C223} &0936+361&0.1368  &  16.0  & 0.74  &  113  &306&1007&$\circ\circ$&2&8.4&N  \\
\Ss{3C223.1} &0938+399&0.1075  &   8.1  & 0.73  &   35  &140& 376&$\circ$&3&8.4&N  \\
\Ss{3C227} &0945+076&0.0861  &  33.1  & 0.70  &   89  &230& 510&&3&8.4&B  \\
\bf\Ss{3C234} &0958+290&0.1848  &  34.2  & 0.86  &  466  &112& 469 &$\bullet$&1&8.4&N \\
\Ss{3C277.3} &1251+278&0.0857  &   9.8  & 0.58  &   26  & 49.0& 108 &$\bullet$&(14)&--&E \\
\bf\Ss{3C284} &1308+277&0.2394  &  12.3  & 0.95  &  299  &178& 904 &&1&8.1&N \\
\bf\Ss{3C285} &1319+428&0.0794  &  12.3  & 0.95  &   29  &180& 371 &$\bullet\circ$&7&4.9&N \\
\bf\Ss{3C300} &1420+198& 0.272  &  19.5  & 0.78  &  604  &100& 561 &$\bullet$&1&8.1&N \\
\bf\Ss{3C303} &1441+522& 0.141  &  12.2  & 0.76  &   92  & 47.0& 159 &$\bullet$&9&1.5&B \\
\bf\Ss{3C319} &1522+546& 0.192  &  16.7  & 0.90  &  249  &105& 453&&1&8.4&E  \\
\bf\Ss{3C321} &1529+242& 0.096  &  14.7  & 0.60  &   49  &307& 748&$\bullet$&(2)&--&N  \\
\Ss{3C327} &1559+021&0.1039  &  38.5  & 0.64  &  152  &302& 788&$\circ\circ$&2&8.4&N  \\
\bf\Ss{3C349} &1658+471& 0.205  &  14.5  & 0.74  &  242  & 85.9& 389 &$\circ$&1&8.4&N \\
\Ss{3C353} &1717+009&0.0304  & 257.2  & 0.74  &   83  &284& 240 &$\bullet\bullet$&8&8.4&E \\
\bf\Ss{3C381} &1832+474&0.1605  &  18.1  & 0.81  &  181  & 73.2& 274&&1&8.4&B  \\
\bf\Ss{3C382} &1833+326&0.0578  &  21.7  & 0.59  &   26  &185& 286 &$\bullet$&3&8.4&B \\
\bf\Ss{3C388} &1842+455&0.0908  &  26.8  & 0.70  &   81  & 50.0& 116&$\bullet\circ$&5&4.9&E  \\
\bf\Ss{3C390.3} &1845+797&0.0561  &  51.8  & 0.75  &   58  &229& 345&$\bullet\circ$&4&8.4&B  \\
\bf\Ss{3C401} &1939+605& 0.201  &  22.8  & 0.71  &  362  & 23.6& 105&$\bullet$&1&8.4&E  \\
\Ss{3C403} &1949+023& 0.059  &  28.3  & 0.74  &   35  &230& 362&$\bullet$&3&8.4&N  \\
\Ss{3C405} &1957+405&0.0565  &9660.0  & 0.74  &11001  &130& 197&$\bullet\bullet$&6&4.5&N  \\
\Ss{3C424} &2045+068& 0.127  &  15.9  & 0.88  &   98  & 35.0& 108 &$\bullet$&3&8.4&E \\
\Ss{3C430} &2117+605&0.0541  &  36.7  & 0.75  &   38  & 90.0& 131&&(13)&--&E  \\
\bf\Ss{3C433} &2121+248&0.1016  &  61.3  & 0.75  &  233  & 68.0& 174& $\bullet$&3&8.4&N \\
\bf\Ss{3C436} &2141+279&0.2145  &  19.4  & 0.86  &  365  &109& 511&$\bullet$&1&8.4&N  \\
\bf\Ss{3C438} &2153+377& 0.290  &  48.7  & 0.88  & 1783  & 22.6& 131&$\bullet\bullet$&1&8.4&E  \\
\Ss{3C445} &2221$-$021&0.0562  &  27.0  & 0.76  &   30  &570& 859 &$\bullet\circ$&2&8.4&B \\
\bf\Ss{3C452} &2243+394&0.0811  &  59.3  & 0.78  &  142  &280& 588 &$\bullet\bullet$&3&8.4&N \\
\end{tabular}
\end{center}
\parbox{\linewidth}{Objects whose names are in bold face are drawn
from the complete sample of LRL. Column 4 gives the 178-MHz flux
density of the source. Column 5 gives the low-frequency (178--750 MHz)
spectral index. Column 6 gives the luminosity at 178 MHz; the units
are $10^{24}$ W Hz$^{-1}$ sr$^{-1}$. Column 7 gives largest angular
size, measured from the best available maps. Column 9 gives
information on jet detection. A
filled circle indicates a definite jet and an open circle a possible
jet. Column 10 gives references to maps used, as follows: (1) H97 and
references therein. (2) L97. (3) B92; Black (1992). (4) Dennett-Thorpe
(1996). (5) Roettiger \etal\ (1994). (6) Perley, Dreher \& Cowan
(1984) (7) van Breugel \& Dey (1993). (8) Swain, Bridle \& Baum
(1996). (9) Leahy \& Perley (1991). (10) Clarke \etal\ (1992). (11)
Leahy, in preparation. (12) Alexander (1985). (13) Spangler, Myers \&
Pogge (1984). (14) van Breugel \etal\ (1985). References in
parentheses are to papers used to classify the jets of objects that
were not included in the analysis. Column 11 gives the observing
frequency of the map used.  Column 12 gives the emission line type of
the source: `E' indicates a LERG, `N' a NLRG and `B' a BLRG; '?'
indicates an unclassified source. These mostly reflect classifications
in LRL, supplemented by references quoted in H97, or from the
spectrophotometry of Laing and co-workers (L94: Laing, private
communication) for sources in LRL. The non-LRL sources have been
classified from the following literature: \Ss{3C15}, \Ss{3C105},
\Ss{3C403}, Tadhunter \etal (1993); \Ss{3C111}, Sargent (1977);
\Ss{3C135}, \Ss{3C445}, Eracleous \& Halpern (1994); \Ss{3C227},
\Ss{3C327}, \Ss{3C353}, Simpson \etal\ (1996); \Ss{3C197.1},
\Ss{3C424} (tentatively), \Ss{3C430}, Smith, Spinrad \& Smith (1976);
\Ss{3C223.1}, Cohen \& Osterbrock (1981); \Ss{3C277.3}, Yee \& Oke
(1978); \Ss{3C405}, Osterbrock \& Miller (1975). \Ss{3C136.1} has
strong emission lines (Smith \etal\ 1976) but is not classified as
broad or narrow-line in the literature.}
\end{table*}

\begin{table*}
\caption{Basic measured quantities}
\label{numbers}
\begin{center}
\begin{tabular}{lld{3}d{3}d{5}d{5}}
Source&Frequency&\multicolumn{1}{l}{Total flux
density}&\multicolumn{1}{l}{Error}&\multicolumn{1}{l}{Core flux density}&\multicolumn{1}{l}{Error}\\
&\multicolumn{1}{l}{(GHz)}&\multicolumn{1}{r}{(Jy)}&&\multicolumn{1}{r}{(mJy)}\\
\hline
\Ss{3C15}&8.35&1.00&&27.99&0.05\\
\Ss{3C20}&8.44&2.29&&3.32&0.06\\
\Ss{3C33.1}&1.53&3.02&&20.40&0.07\\
\Ss{3C79}&8.44&0.694&&6.04&0.01\\
\Ss{3C98}&8.35&3.08&0.07&6.1&0.1\\
\Ss{3C105}&8.35&1.68&&18.9&0.5\\
\Ss{4C14.11}&8.44&0.500&&29.69&0.03\\
\Ss{3C111}&8.35&4.8&0.2&1276&1\\
\Ss{3C123}&8.44&9.44&&108.9&0.3\\
\Ss{3C132}&8.44&0.674&&4.1&0.2\\
\Ss{3C135}&8.35&0.520&&1.0&0.2\\
\Ss{3C136.1}&8.35&1.00&0.05&1.53&0.03\\
\Ss{3C153}&8.44&0.712&&<0.5&\\
\Ss{3C171}&8.06&0.690&&2.0&0.1\\
\Ss{3C173.1}&8.44&0.461&&9.64&0.02\\
\Ss{3C184.1}&8.35&0.785&&6.0&0.5\\
\Ss{3C192}&8.35&1.38&0.08&4.0&0.2\\
\Ss{3C197.1}&8.35&0.320&&6.0&0.1\\
\Ss{3C219}&4.87&2.27&0.06&51.6&0.1\\
\Ss{3C223}&8.35&0.89&0.05&8.5&0.2\\
\Ss{3C223.1}&8.35&0.53&0.01&6.4&0.4\\
\Ss{3C227}&8.35&2.05&0.04&13.2&0.6\\
\Ss{3C234}&8.44&0.919&&34.46&0.04\\
\Ss{3C284}&8.06&0.340&&2.79&0.02\\
\Ss{3C285}&4.86&0.740&&6.8&0.4\\
\Ss{3C300}&8.06&0.645&&6.2&0.1\\
\Ss{3C303}&1.48&2.45&&106.6&0.3\\
\Ss{3C319}&8.44&0.362&&<0.3&\\
\Ss{3C327}&8.35&2.01&0.05&25&1\\
\Ss{3C349}&8.44&0.723&&24.21&0.02\\
\Ss{3C353}&8.44&14.1&&151.0&0.2\\
\Ss{3C381}&8.44&0.906&&4.7&0.1\\
\Ss{3C382}&8.35&1.30&&251.2&0.1\\
\Ss{3C388}&4.87&1.80&&57.9&0.1\\
\Ss{3C390.3}&8.35&2.8&0.1&733&5\\
\Ss{3C401}&8.44&0.844&&28.54&0.03\\
\Ss{3C403}&8.35&1.50&&7.1&0.2\\
\Ss{3C405}&4.53&415&&776&3\\
\Ss{3C424}&8.35&0.357&&7.0&0.3\\
\Ss{3C433}&8.35&2.08&&1.2&0.3\\
\Ss{3C436}&8.44&0.592&&17.90&0.02\\
\Ss{3C438}&8.44&0.780&&16.2&0.1\\
\Ss{3C445}&8.40&1.34&0.08&83.9&0.4\\
\Ss{3C452}&8.35&2.14&&125.8&0.3\\
\end{tabular}
\end{center}
\parbox{\linewidth}{
The total flux densities for \Ss{3C98}, \Ss{3C136.1}, \Ss{3C192},
\Ss{3C227}, \Ss{3C327} and \Ss{3C445} are taken from the single-dish
measurements of Stull (1971), as these sources were all seriously
undersampled by the VLA observations. These numbers are to be treated
with caution. Stull's flux scale is not consistent with that of Baars
\etal\ (1977), but the systematic error is much less than 1 per
cent. The random error due to the calibration process is tabulated
above and is as given by Stull. In addition, the flux densities have
not been corrected from 8.0 GHz to the VLA frequency, which makes them
systematically high by up to 5 per cent. The 5-GHz flux density of
\Ss{3C219} is from Laing \& Peacock (1980). The flux densities for
\Ss{3C223} and \Ss{3C390.3} were interpolated from other frequencies,
and the errors assigned to them are estimates of the error in the
interpolation. Elsewhere errors are only assigned where the
integration was problematic because of off-source noise, but the VLA
calibration errors apply in addition to these and to any source with
no error assigned. Noise or mild undersampling means that the flux densities
of \Ss{3C105}, \Ss{3C111}, \Ss{3C132}, \Ss{3C171}, \Ss{3C197.1},
\Ss{3C284} and \Ss{3C382} may be too low. The flux density of \Ss{3C303} is
taken from Leahy \& Perley (1991).  The core flux densities of 
\Ss{3C79}, \Ss{4C14.11}, \Ss{3C111} and \Ss{3C390.3} are variable (see H97, L97 and
Leahy \& Perley 1995). We took the southern component of 3C424's core
to be the true core ({\it pace} B92).}
\end{table*}

\begin{table*}
\caption{Flux densities and sizes of lobes}
\label{lobes}
\begin{center}
\begin{tabular}{ll@{\hspace{7mm}}d{3}d{3}d{1}@{\hspace{7mm}}d{3}d{3}d{1}}
&&\multicolumn{3}{c}{North lobe}&\multicolumn{3}{c}{South lobe}\\
Source&Frequency&\multicolumn{1}{l}{Flux
density}&\multicolumn{1}{l}{Error}&\multicolumn{1}{l}{Length}&\multicolumn{1}{l}{Flux
density}&\multicolumn{1}{l}{Error}&\multicolumn{1}{l}{Length}\\
&(GHz)&\multicolumn{1}{l}{(Jy)}&&\multicolumn{1}{l}{(arcsec)}&\multicolumn{1}{l}{(Jy)}&&\multicolumn{1}{l}{(arcsec)}\\
\hline
\Ss{3C15}&8.35&0.54&0.01&24.6&0.41&0.01&23.6\\
\Ss{3C20}&8.44&1.23&0.02&27.2&1.06&0.02&28.7\\
\Ss{3C33.1}&1.53&1.73&0.02&87.9&1.28&0.02&152\\
\Ss{3C79}&8.44&0.386&&38.3&0.302&&50.8\\
\Ss{3C98}&8.35&1.67&&138&1.44&&178\\
\Ss{3C105}&8.35&0.356&&162&1.31&&171\\
\Ss{4C14.11}&8.44&0.231&0.005&58.6&0.239&0.005&57.5\\
\Ss{3C111}&8.35&-&&123&-&&93.9\\
\Ss{3C123}&8.44&3.31&&19.2&6.02&&17.6\\
\Ss{3C132}&8.44&0.273&0.005&11.7&0.397&0.005&11.0\\
\Ss{3C135}&8.35&0.186&0.005&78.8&0.337&0.005&51.7\\
\Ss{3C136.1}&8.35&-&&191&-&&267\\
\Ss{3C153}&8.44&0.355&0.002&4.3&0.356&0.002&4.8\\
\Ss{3C171}&8.06&0.386&&24.4&0.302&&12.1\\
\Ss{3C173.1}&8.44&0.201&&27.2&0.250&&33.4\\
\Ss{3C184.1}&8.35&0.397&0.001&107&0.382&0.001&82.0\\
\Ss{3C192}&8.35&-&&109&-&&93.0\\
\Ss{3C197.1}&8.35&0.147&0.005&16.0&0.170&0.005&11.6\\
\Ss{3C219}&4.87&-&&97.8&-&&92.1\\
\Ss{3C223}&8.35&0.440&&156&0.320&&152\\
\Ss{3C223.1}&8.35&0.180&&81.0&0.215&&60.0\\
\Ss{3C227}&8.35&1.03&&113&0.660&&121\\
\Ss{3C234}&8.44&0.334&0.005&64.6&0.551&0.005&48.0\\
\Ss{3C284}&8.06&0.152&0.003&106&0.185&0.003&72.6\\
\Ss{3C285}&4.86&0.408&&87.0&0.325&&103\\
\Ss{3C300}&8.06&0.154&0.005&70.2&0.485&0.005&31.8\\
\Ss{3C303}&1.48&1.90&0.02&27.8&0.40&0.01&20.8\\
\Ss{3C319}&8.44&0.25&0.01&48.9&0.12&0.01&57.9\\
\Ss{3C327}&8.35&-&&199&-&&108\\
\Ss{3C349}&8.44&0.269&0.008&41.7&0.430&0.008&44.2\\
\Ss{3C353}&8.44&8.5&0.1&142&5.5&0.1&142\\
\Ss{3C381}&8.44&0.570&&33.2&0.331&&40.0\\
\Ss{3C382}&8.35&0.790&&87.9&0.580&&91.0\\
\Ss{3C388}&4.87&0.869&&27.9&0.873&&22.5\\
\Ss{3C390.3}&8.35&-&&132&-&&92.0\\
\Ss{3C401}&8.44&0.348&0.005&10.7&0.467&0.005&12.9\\
\Ss{3C403}&8.35&-&&95.0&-&&116\\
\Ss{3C405}&4.53&205&&70.6&209&&60.6\\
\Ss{3C424}&8.35&0.120&0.005&13.0&0.230&0.005&21.0\\
\Ss{3C433}&8.35&0.237&0.005&41.0&1.84&0.01&32.5\\
\Ss{3C436}&8.44&0.27&0.01&59.2&0.30&0.01&49.6\\
\Ss{3C438}&8.44&0.368&0.005&11.9&0.397&0.005&10.8\\
\Ss{3C445}&8.40&-&&299&-&&280\\
\Ss{3C452}&8.35&0.940&&141&1.07&&135\\

\end{tabular}
\end{center}
\parbox{\linewidth}{Flux densities are listed only where maps were available to
us that appeared to reproduce the large-scale structure well. Errors
have been assigned where there was some difficulty in separating the
north and south lobes, and so where some flux might be assigned to
either; the errors are rough estimates of the variation over multiple
attempts at integration. These are in addition to the errors due to
VLA flux calibration, which apply to all the measurements. Lengths are
measured from the core to the most distant visible region of
emission. In several cases these lengths are measured substantially
away from the source axis: these, with their on-axis lengths in
parentheses, are \Ss{3C105}N (156 arcsec), \Ss{3C171}N (5.0) and S
(4.8), \Ss{3C197.1}N (7), \Ss{3C300}S (30), \Ss{3C403}N (55.5) and S
(52.8) and \Ss{3C433}N (24.6).}
\end{table*}

\begin{table*}
\caption{Flux densities and lengths of jets}
\label{jet_tab}
\begin{center}
\begin{tabular}{ll@{\hspace{5mm}}d{2}d{2}d{2}@{\hspace{5mm}}d{2}d{2}d{2}}
&&\multicolumn{3}{c}{North jet}&\multicolumn{3}{c}{South jet}\\
Source&Freq.&\multicolumn{1}{c}{Flux
density}&\multicolumn{1}{c}{Error}&\multicolumn{1}{c}{Length}&\multicolumn{1}{c}{Flux
density}&\multicolumn{1}{c}{Error}&\multicolumn{1}{c}{Length}\\
&(GHz)&\multicolumn{1}{c}{(mJy)}&&&\multicolumn{1}{c}{(mJy)}\\
\hline
\Ss{3C15}&8.35&210&10&9.50&3.6&0.9&6.20\\
\Ss{3C20}&8.44&23&6&18.3&^*3.7&0.9&6.80\\
\Ss{3C33.1}&1.53&<56&&&170&10&78.0\\
\Ss{3C79}&8.44&<3.8&&&<1.3&&\\
\Ss{3C98}&8.35&90&20&121&<0.79&&\\
\Ss{3C105}&8.35&<8.3&&&^*40&10&17.2\\
\Ss{4C14.11}&8.44&<2.5&&&1.9&0.3&2.80\\
\Ss{3C111}&8.35&100&10&114&<2.8&&\\
\Ss{3C123}&8.44&<2.1&&&<30&&\\
\Ss{3C132}&8.44&<7.2&&&^*3&2&5.97\\
\Ss{3C135}&8.35&<3&&&3.1&0.4&21.4\\
\Ss{3C136.1}&8.35&<2.6&&&<2.7&&\\
\Ss{3C153}&8.44&^*1.9&0.6&1.20&^*13&2&1.40\\
\Ss{3C171}&8.06&15&1&3.80&6.3&0.9&3.50\\
\Ss{3C173.1}&8.44&2.4&0.8&10.2&<2.4&&\\
\Ss{3C184.1}&8.35&^*14&3&32.2&<1.3&&\\
\Ss{3C192}&8.35&<3.2&&&^*8&1&13.9\\
\Ss{3C197.1}&8.35&<1.9&&&<0.85&&\\
\Ss{3C219}&4.87&^*2.09&0.09&1.40&56.1&0.5&16.5\\
\Ss{3C223}&8.35&^*8&2&29.4&^*5&9&42.0\\
\Ss{3C223.1}&8.35&^*2.5&0.7&19.2&<7.7&&\\
\Ss{3C227}&8.35&<15&&&<27&&\\
\Ss{3C234}&8.44&3&2&6.20&<2.3&&\\
\Ss{3C284}&8.06&<2.8&&&<6.6&&\\
\Ss{3C285}&4.86&36&7&74.6&<2.1&&\\
\Ss{3C300}&8.06&38&4&62.6&<1.7&&\\
\Ss{3C303}&1.48&66&4&14.2&<7.4&&\\
\Ss{3C319}&8.44&<0.12&&&<1.8&&\\
\Ss{3C327}&8.35&^*11&3&107&^*11&3&60.4\\
\Ss{3C349}&8.44&<2.1&&&^*0.24&0.04&1.10\\
\Ss{3C353}&8.44&90&10&71.0&30&8&37.0\\
\Ss{3C381}&8.44&<2.9&&&<1.3&&\\
\Ss{3C382}&8.35&31&2&77.9&<0.14&&\\
\Ss{3C388}&4.87&^*10&2&4.58&49&8&12.7\\
\Ss{3C390.3}&8.35&34&5&81.3&^*2&1&21.0\\
\Ss{3C401}&8.44&<4&&&114&9&5.85\\
\Ss{3C403}&8.35&7.8&0.6&24.0&-&&-\\
\Ss{3C405}&4.53&2300&300&50.3&380&40&24.1\\
\Ss{3C424}&8.35&<11&&&15&1&3.04\\
\Ss{3C433}&8.35&47&9&24.1&<1.4&&\\
\Ss{3C436}&8.44&<0.51&&&19&3&38.8\\
\Ss{3C438}&8.44&43&9&8.90&40&3&7.40\\
\Ss{3C445}&8.40&^*1.3&0.3&5.60&10.3&0.7&60.0\\
\Ss{3C452}&8.35&16&3&95.4&31&7&108\\
\end{tabular}
\end{center}
\parbox{\linewidth}{Flux densities of jets classed as `possible' in
Table \ref{combsamp} are marked with an asterisk.}
\end{table*}
\begin{table*}

\caption{Flux densities and lengths of straight jets}
\label{stjet_tab}
\begin{center}
\begin{tabular}{lld{3}d{3}d{3}d{3}d{2}}
&&\multicolumn{2}{c}{North straight jet}&\multicolumn{2}{c}{South
straight jet}\\
Source&Freq.&\multicolumn{1}{c}{Flux
density}&\multicolumn{1}{c}{Error}&\multicolumn{1}{c}{Flux density}&\multicolumn{1}{c}{Error}&\multicolumn{1}{c}{Length}\\
&(GHz)&\multicolumn{1}{c}{(mJy)}&&\multicolumn{1}{c}{(mJy)}\\
\hline
\Ss{3C15}&8.35&96&3&<1.4&&4.17\\
\Ss{3C20}&8.44&7&5&^*<9&&12.4\\
\Ss{3C33.1}&1.53&<37&&27&4&51.0\\
\Ss{3C79}&8.44&-&&-&\\
\Ss{3C98}&8.35&50&20&<13&&101\\
\Ss{3C105}&8.35&-&&-&\\
\Ss{4C14.11}&8.44&<0.45&&1.14&0.04&15.0\\
\Ss{3C111}&8.35&90&20&<58&&101\\
\Ss{3C123}&8.44&-&&-&\\
\Ss{3C132}&8.44&<13&&^*2&3&5.00\\
\Ss{3C135}&8.35&<1.6&&10&4&30.2\\
\Ss{3C136.1}&8.35&-&&-&\\
\Ss{3C153}&8.44&^*8&1&^*13&5&1.70\\
\Ss{3C171}&8.06&6.0&0.7&6.3&0.8&3.07\\
\Ss{3C173.1}&8.44&2.1&0.9&<0.29&&11.2\\
\Ss{3C184.1}&8.35&-&&-&\\
\Ss{3C192}&8.35&-&&-&\\
\Ss{3C197.1}&8.35&-&&-&\\
\Ss{3C219}&4.87&^*56.5&0.3&2.1&0.1&17.1\\
\Ss{3C223}&8.35&^*11&4&^*<5.8&&34.1\\
\Ss{3C223.1}&8.35&^*3&1&<21&&21.9\\
\Ss{3C227}&8.35&-&&-&\\
\Ss{3C234}&8.44&10&8&<19&&59.1\\
\Ss{3C284}&8.06&-&&-&\\
\Ss{3C285}&4.86&19&2&<14&&51.6\\
\Ss{3C300}&8.06&2.4&0.2&<0.23&&4.70\\
\Ss{3C303}&1.48&63&5&<13&&12.3\\
\Ss{3C319}&8.44&-&&-&\\
\Ss{3C327}&8.35&^*16&6&^*<140&&119\\
\Ss{3C349}&8.44&<0.053&&^*0.31&0.04&1.27\\
\Ss{3C353}&8.44&70&10&<34&&66.5\\
\Ss{3C381}&8.44&-&&-&\\
\Ss{3C382}&8.35&14&1&<1.1&&48.3\\
\Ss{3C388}&4.87&^*30&5&34&6&8.40\\
\Ss{3C390.3}&8.35&20&10&^*<650&&88.6\\
\Ss{3C401}&8.44&<5.7&&33.8&0.4&5.67\\
\Ss{3C403}&8.35&6&1&<3.8&&19.6\\
\Ss{3C405}&4.53&1200&600&500&800&40.8\\
\Ss{3C424}&8.35&<8.4&&16.7&0.9&2.99\\
\Ss{3C433}&8.35&9.8&0.1&<8.3&&4.43\\
\Ss{3C436}&8.44&<0.27&&3.8&0.8&20.2\\
\Ss{3C438}&8.44&40&4&<9.8&&8.67\\
\Ss{3C445}&8.40&^*<1.9&&14&3&78.1\\
\Ss{3C452}&8.35&9&2&13&2&46.4\\
\end{tabular}
\end{center}
\parbox{\linewidth}{\small Flux densities of jets classed as
`possible' in Table \ref{combsamp} are
marked with an asterisk.}
\end{table*}

\begin{table*}
\caption{Flux densities and sizes of hot spots}
\label{bhot_tab}
\begin{center}
\begin{tabular}{lld{1}d{1}d{2}d{2}d{2}d{1}d{1}d{2}d{2}d{2}}
&&\multicolumn{5}{c}{North hot spot}&\multicolumn{5}{c}{South hot spot}\\
Source&Freq.&\multicolumn{1}{c}{Flux density}&\multicolumn{1}{c}{Error}&\multicolumn{1}{c}{$\theta_{\rm
maj}$}&\multicolumn{1}{c}{$\theta_{\rm
min}$}&\multicolumn{1}{c}{Dist.}&\multicolumn{1}{c}{Flux density}&\multicolumn{1}{c}{Error}&\multicolumn{1}{l}{$\theta_{\rm maj}$}&\multicolumn{1}{l}{$\theta_{\rm min}$}&\multicolumn{1}{c}{Dist.}\\
&(GHz)&\multicolumn{1}{c}{(mJy)}&&&&&\multicolumn{1}{c}{(mJy)}\\
\hline
\Ss{3C15}&8.35&-&&-&-&-&^*5&1&1.5&1.0&18.2\\
\Ss{3C20}&8.44&160&10&0.21&0.16&24.1&89&2&0.22&0.18&23.3\\
\Ss{3C79}&8.44&4&1&0.21&0.08&36.8&15&2&0.53&0.34&50.4\\
\Ss{3C98}&8.35&^*40&10&4.1&3.4&132&27&5&3.9&2.0&155\\
\Ss{3C105}&8.35&3.2&0.6&1.7&0.80&155&110&5&0.28&0.20&171\\
\Ss{4C14.11}&8.44&1.7&0.5&1.2&0.10&58.1&3.3&0.1&0.29&0.09&36.4\\
\Ss{3C111}&8.35&300&10&1.3&0.70&123&76&1&1.9&1.1&74.1\\
\Ss{3C123}&8.44&21&2&0.15&0.08&7.3&162&5&0.11&0.08&7.9\\
\Ss{3C132}&8.44&53&1&0.19&0.15&11.3&22&4&0.40&0.25&10.9\\
\Ss{3C135}&8.35&1.7&0.2&0.55&0.25&72.3&^*77&5&4.6&1.7&45.0\\
\Ss{3C136.1}&8.35&^*56&8&20&10&173&^*23&5&12&12&255\\
\Ss{3C153}&8.44&133&2&0.19&0.08&2.4&87&1&0.11&0.07&4.8\\
\Ss{3C171}&8.06&120&10&0.27&0.07&4.9&95&2&0.29&0.19&4.6\\
\Ss{3C173.1}&8.44&8.3&0.3&0.47&0.18&26.4&10&2&0.50&0.25&31.9\\
\Ss{3C184.1}&8.35&^*7&2&2.6&1.0&103&19&1&0.95&0.50&78.2\\
\Ss{3C192}&8.35&^*80&20&4.0&3.0&103&2.0&0.2&1.2&0.90&88.5\\
\Ss{3C197.1}&8.35&3.4&0.1&0.33&0.30&6.8&8.5&0.5&0.75&0.35&9.6\\
\Ss{3C219}&4.87&2.9&0.2&0.99&0.36&72.6&76&1&3.2&1.6&73.1\\
\Ss{3C223}&8.35&10&1&3.5&1.3&140&^*6&2&6.0&2.0&147\\
\Ss{3C223.1}&8.35&^*14&4&2.0&0.80&40.3&13&1&0.90&0.40&38.5\\
\Ss{3C227}&8.35&17&1&0.90&0.60&108&^*7&1&1.0&0.50&109\\
\Ss{3C234}&8.44&55.8&0.1&0.51&0.20&63.9&50&10&0.73&0.35&47.4\\
\Ss{3C284}&8.06&^*6&2&1.5&0.75&104&27&2&0.76&0.57&72.4\\
\Ss{3C285}&4.86&^*4&1&3.0&2.0&78.8&^*4&1&9.0&7.0&92.5\\
\Ss{3C300}&8.06&^*1.0&0.5&0.30&0.30&69.7&29&1&0.44&0.26&29.3\\
\Ss{3C303}&1.48&650&10&<1.7&1.1&16.9&5.0&0.5&0.90&0.50&16.9\\
\Ss{3C319}&8.44&19&3&1.4&0.90&47.7&-&&-&-&-\\
\Ss{3C327}&8.35&3.5&0.2&0.40&0.30&182&20&1&1.0&0.35&99.1\\
\Ss{3C349}&8.44&^*4&1&0.50&0.30&40.3&90&5&0.65&0.47&42.9\\
\Ss{3C353}&8.44&^*70&10&4.5&2.5&94.4&63&2&3.2&1.8&123\\
\Ss{3C381}&8.44&9&1&0.20&0.17&32.5&^*7&2&0.80&0.40&35.5\\
\Ss{3C382}&8.35&45&5&2.2&2.0&86.5&^*8&2&2.3&2.1&82.4\\
\Ss{3C388}&4.87&40&10&2.8&1.6&15.9&55&2&1.2&0.90&16.4\\
\Ss{3C390.3}&8.35&67&2&2.5&1.2&104&450&20&4.1&2.0&88.4\\
\Ss{3C401}&8.44&^*5&2&0.50&0.50&8.8&^*4&2&0.60&0.20&12.3\\
\Ss{3C403}&8.35&30&1&0.40&0.21&28.5&^*20&10&4.0&0.90&47.8\\
\Ss{3C405}&4.53&3060&20&0.61&0.35&63.2&2320&50&1.0&0.47&53.1\\
\Ss{3C424}&8.35&2.7&0.2&0.20&0.15&8.7&24&2&0.50&0.20&4.2\\
\Ss{3C433}&8.35&-&&-&-&-&^*6&2&0.65&0.55&0.0\\
\Ss{3C436}&8.44&^*2&1&2.2&1.0&57.4&11&1&0.34&0.21&43.5\\
\Ss{3C438}&8.44&^*4&2&0.30&0.10&11.4&^*2&1&0.30&0.10&8.9\\
\Ss{3C445}&8.40&43&2&2.2&1.0&291&60&10&3.7&1.4&275\\
\Ss{3C452}&8.35&^*20&10&3.5&1.2&130&31&1&1.0&0.84&126\\
\end{tabular}
\end{center}
\parbox{\linewidth}{$\theta_{\rm maj}$, $\theta_{\rm min}$ and `Dist.' are in arcseconds. `Dist.' is the
distance between the hot spot and the core. Flux densities marked with
an asterisk were measured by integration, and their associated major
and minor axes ($\theta_{\rm maj}$ and $\theta_{\rm min}$) by
estimation from the maps rather than by using Gaussian fitting. In several
cases it was not clear which object was the primary; in these cases
fits were made to each candidate component and the results for the
most compact components were used here. The relevant objects are
\Ss{3C173.1}N (N4 was used rather than N3), \Ss{3C227}N (F1a of B92
was used) \Ss{3C285}N (southern candidate object was preferred),
\Ss{3C300}S (E2 was used rather than E3) and \Ss{3C403}N (F6 of B92
was taken to be the primary rather than F1). \Ss{3C33.1} is omitted,
as the maps available to us were not high enough in resolution to
allow a measurement.}
\end{table*}

\begin{table*}
\caption{Median values of important quantities}
\label{median}
\begin{center}
\begin{tabular}{llllllll}
Quantity&All radio galaxies&NLRG&BLRG&NLRG and BLRG&LERG&B94 quasars\\
\hline
$z$&0.129 (0.182)&0.123 (0.179)&0.086 (0.150)&0.118 (0.167)&0.197 (0.210)&0.768\\
178-MHz luminosity&115 (237)&151 (241)&89 (134)&129
(181)&103 (268)&6600\\
 ($10^{24}$ W Hz$^{-1}$ sr$^{-1}$)\\
Linear size (kpc)&367 (380)&409 (468)&344 (315)&406 (404)&130
(179)&418\\
Hot spot size (kpc)&2.42 (2.42)&2.38 (2.57)&3.32 (2.89)&2.49
(2.92)&2.15 (2.15)&--\\
Core prominence&0.012 (0.014)&0.0087 (0.0087)&0.067 (0.12)&0.011
(0.010)&0.021 (0.023)&0.062\\
Straight jet prominence&0.0064 (0.0067)&0.0055 (0.0057)&0.011
(0.011)&0.0067 (0.0069)&0.0049 (0.0042)&0.0088\\
Total jet prominence&0.014 (0.018)&0.013 (0.018)&0.026 (0.029)&0.017
(0.019)&0.0063 (0.0051)&0.053\\
\end{tabular}
\end{center}
\parbox{\linewidth}{Values in parentheses are drawn from the
sub-sample of objects taken from LRL (which constitute a flux-limited
sample).}
\end{table*}

\begin{table*}
\caption{Summary of trends and correlations}
\label{summary}
\begin{center}
\begin{tabular}{ll}
Proposition tested&Significant?\\
\hline
LERG are smaller than NLRG and BLRG&Y (99.9 per cent)\\
BLRG are smaller than NLRG&N\\[3pt]
Hot spot size is correlated with total linear size&Y (99.9 per cent)\\
The brighter jet points towards the more compact hot spot&N\\
The hot spot on the jetted size is less recessed&Y? (90 per cent)\\
The brighter jet points towards the brighter hot spot&Y? (90 per
cent)\\[3pt]
LERG are less symmetrical than NLRG&N\\
BLRG are more symmetrical than NLRG&Y (97 per cent)\\[3pt]
Jet prominence depends on luminosity&N\\
Jet detection fraction depends on emission line class&N\\
BLRG straight jets are more prominent than NLRG straight jets&Y (95 per cent)\\
B94 quasar total jets are more prominent than BLRG total jets&Y (95 per
cent)\\
B94 quasar straight jets are more prominent than BLRG straight
jets&N\\
Jets are preferentially detected in shorter objects&Y (90 per cent)\\
Brighter jets lie in longer lobes&N\\
Brighter jets lie in brighter lobes&N\\
Jet length is correlated with total linear size&Y (99 per cent)\\[3pt]
BLRG cores are more prominent than NLRG cores&Y (99 per cent)\\
LERG cores and NLRG/BLRG cores are drawn from distributions with
different medians&N\\
LERG cores and NLRG/BLRG cores are drawn from different
distributions&Y? (90 per cent)\\
B94 quasar cores are more prominent than BLRG cores&N\\
NLRG core prominence is correlated with total linear size&Y (99 per
cent)\\
Straight jet prominence is correlated with core prominence&Y (95 per
cent)\\
\end{tabular}
\end{center}
\parbox{\linewidth}{Trends and correlations are considered significant (Y) if the probability of the observed result under the
null hypothesis is $\le 5$ per cent,  marginally
significant (Y?) if it is $\le 10$ per cent
and not significant (N) otherwise.}
\end{table*}

\clearpage

\end{document}